# Near-visible low power tuning of nematic-liquid crystal integrated silicon nitride ring resonator

Jayita Dutta, [*,†] Antonio Ferraro, [§] Arnab Manna,[¶] Rui Chen,[†] Alfredo Pane[§], Giuseppe Emanuele Lio,[||] Roberto Caputo,[‡, §] and Arka Majumdar[*,†,¶]

† Electrical and Computer Engineering, University of Washington, Seattle, WA, 98195, USA.

,§ Consiglio Nazionale delle Ricerche - Istituto di Nanotecnologia CNR-Nanotec, Rende (CS), 87036 Italy

¶ Department of Physics, University of Washington, Seattle, WA, 98195, USA.

|| Istituto di Nanoscienze CNR-NANO, Consiglio Nazionale delle Ricerche, Pisa, 56127, Italy

‡ University of Calabria, I-87036 Rende (CS), Italy.

E-mail: jayitad@uw.edu; arka@uw.edu

## Abstract

The development of compact, low-power, and high-performance integrated photonic phase shifters is critical for advancing emerging technologies such as light detection and ranging (LiDAR), optical information processing and quantum applications. Liquid crystal (LC)-based phase shifters offer a promising solution thanks to their large refractive index contrast and their low power consumption. However, it remains challenging to incorporate LCs into integrated photonics and the operating wavelength has been limited to near infrared. Here, we demonstrate a liquid-crystal-based phase shifter operating at 780 nm, a relevant wavelength for cold atom-based quantum applications, by incorporating nematic LCs (E7) into a silicon nitride (SiN) ring resonator. Our device achieves ~$2\pi$ phase modulation with very low power of 2.1 nW and low driving voltages of ~ 7 V with exceptionally low $V\pi \cdot L$ ($half\ wave\ voltage\ .\ length$) value of 0.014 V·cm, enabling precise control over light propagation in a compact footprint. This work marks a significant step toward realizing low-power, compact, and efficient LC integrated photonic circuits based on SiN platform for next-generation LiDAR and quantum optical systems.

## Introduction

The advancement of compact, low-power, and high-performance integrated photonic systems is crucial for enabling next-generation technologies, including quantum computing, Light Detection and Ranging (LiDAR), and optical beam steering[1–5]. In quantum computing, photonic integration facilitates scalable and low-loss quantum information processing by providing efficient light manipulation, on-chip entanglement generation, and precise single-photon detection[6–10]. For LiDAR, integrated photonics enhance system miniaturization while reducing power consumption, improving depth perception, and enabling real-time 3D mapping for autonomous vehicles and robotics[11–19]. Similarly, in optical beam steering, integrated photonic circuits can potentially replace traditional bulky mechanical components with fast, energy-efficient, and highly controllable phase shifters and waveguide arrays, enabling agile and precise beam control for applications such as free-space optical communications, augmented reality, and satellite-based sensing[20–23]. The continuous development of photonic materials, fabrication techniques, and innovative device architectures plays a pivotal role in achieving these performance benchmarks, making integrated photonic systems indispensable for the future of high-speed, energy-efficient, and scalable optical technologies[24–30].

Silicon has conventionally served as the platform in integrated photonics due to its compatibility with mature CMOS fabrication processes, ability to support high index contrast for compact device design, and ease of modulation[4,30–35]. However, silicon absorbs visible and some of the near infrared wavelengths, making it unsuitable for integrated photonic systems operating in the wavelength range of 300 nm to 1000 nm[4,35,36]. Instead, silicon nitride (SiN) exhibits very low optical losses spanning from the visible to the mid-infrared wavelengths, making it preferable for applications in various spectral ranges [37,38]. While SiN offers a wide transparency window, it has a low thermo-optic coefficient (~10 times lower than silicon) and lacks strong electro-optic properties, making integrated phase modulation challenging[39–41]. Conventional SiN-based phase modulators, which rely on thermal tuning, typically span hundreds of microns to several millimeters and consume excessive power, imposing significant limitations on system efficiency and scalability[42–44]. Therefore, compact, low-power SiN phase modulators are a critical advancement to enable efficient integrated photonic systems.

Liquid-crystal (LC)-based phase modulators present a promising solution for integrated photonic systems due to their unique advantages of low power consumption and electrically tunable refractive index[45–48]. In particular, nematic LCs, exhibit strong birefringence at optical wavelengths making them highly attractive for applications in quantum computing, LiDAR, and optical beam steering, where precise and energy-efficient phase modulation is essential[49–55]. In quantum photonics, LC modulators enable dynamic control of optical paths and quantum state encoding without introducing significant thermal effects[56–60]. In LiDAR, LCs offer a means to achieve non-mechanical beam steering, improving system reliability and reducing power demands[61–64]. Similarly, in optical beam steering, their ability to continuously tune the refractive index allows for agile, high-resolution beam deflection without the need for moving parts[65–68].

However, despite these advantages, integrating LC based phase modulators into SiN photonic platforms remains a challenge due to fabrication complexity, limited compatibility with existing photonics foundries, and response time constraints[69]. Overcoming these hurdles requires advancements in material engineering, device design, and hybrid integration techniques to seamlessly incorporate LC modulators into compact, high-performance, integrated SiN platform, ultimately unlocking their full potential for next-generation optical technologies.

Various integrated LC-based devices, including rectangular-waveguide and slot-waveguide phase shifters, as well as ring resonators, have been explored in the past[70–78]. However, most of these implementations have been predominantly based on silicon waveguides and not able to support near visible operation. Yet, the near visible spectrum range holds significant importance in integrated photonics, particularly in quantum applications, due to its strong relevance in atomic physics and quantum information processing[1]. This wavelength corresponds to the D2 transition of rubidium (Rb), a commonly used atomic species in quantum technologies, including atomic clocks, magnetometers, and quantum memory systems[79,80]. Thus, integration of liquid crystal into SiN platform with oat 780 nm scan help researchers develop compact, on-chip, low-power, and robust quantum devices, paving the way for scalable quantum networks, quantum sensing, and atom-based quantum computing.

In this work, we developed a method to integrate LC onto SiN waveguides to realize phase shifters. Tested in a ring resonator at a near-visible wavelength of 780 nm, our LC-based phase shifter exhibits an exceptionally low $V\pi \cdot L$ of just 0.014 V·cm, significantly surpassing the performance of current state-of-the-art thermo-optic and electro-optic phase shifters, which typically have $V\pi \cdot L$ values in the range of several V·cm[81–89]. Moreover, we varied the distance between the metal and waveguide edge both in simulation and experiments to study the variation in the phase shift and loss. Both simulation and experimental results suggest that an increase in the distance decreases the phase shift as well as the loss. A metal-waveguide edge separation of 1 $\mu m$ is chosen to achieve a good tradeoff between the phase shift and loss, which shows full-FSR (Free Spectral Range) tuning capabilities with low driving voltage of 7 V and a low power of only ~ 2.1 $n$W. Experiments show ~0.08 dB/$\mu m$ excessive losses due to the LC, which increases to ~0.34 dB/$\mu m$ on applying an electric field of 7 V/$\mu m$ to switch the LC. Regarding the response time, we observed a much faster rise time (~1 ms) compared to the fall time (~4 s), which is attributed to the slow relaxation of LC molecules inside the ring resonator. However, we emphasize that the slow relaxation ensures that the last-written optical state persists for ~4s even after the voltage is removed, providing a quasi-static memory. This non-volatile behavior reduces power consumption, as the system does not require continuous voltage to maintain the optical state[90]. This work marks a significant step toward realizing low-power, compact, efficient LC integrated photonic circuits based on SiN platform for next-generation LiDAR and quantum optical systems.

**Results and Discussion**

We use nematic LC E7 to provide optical modulation for the SiN based device. E7 is a widely used LC mixture known for its excellent electro-optic properties and stability, making it ideal for applications in displays, optical modulators, and tunable photonic devices[91–94]. It is a eutectic mixture of several cyanobiphenyl and terphenyl compounds, giving it a high birefringence of ~0.2 and a relatively low viscosity, which enables relatively faster response times[95,96]. E7 exhibits a broad nematic phase range, typically from approximately -10°C to 60°C, allowing it to function effectively in various environmental conditions[97–99]. Its dielectric anisotropy is positive;

thus, the molecules tend to align parallel to an applied electric field, making it well-suited for electrically tunable devices such as phase shifters and spatial light modulators[100–102]. Additionally, its compatibility with various alignment layers and substrates makes it adaptable for integration into SiN platform[103,104].

The LC phase shifters are constituted of micro ring resonators, where phase shift (loss) is represented by the resonance shift (resonance broadening). **Error! Reference source not found.**a shows the schematic of a SiN micro ring resonator under a 1 $\mu m$ silicon dioxide ($SiO_2$) cladding. A 1 $\mu m$ trench of 40 $\mu m$ length is etched into the $SiO_2$ cladding above the waveguide, which is later filled with LCs. Aluminum electrodes are deposited on both sides of the trench to apply the electric field across the LC. The cross-sectional view of the device is shown in **Error! Reference source not found.**b. The micro ring resonators were fabricated on a standard SiN-on-insulator wafer with a 220 nm SiN layer on 4-$\mu m$ buried $SiO_2$ layer, as shown in Figure 2. The 500-nm-wide waveguides were created via E-beam lithography followed by Fluorine-based dry etching. All the rings have a bus-ring gap of 100 nm to achieve a near-critical coupling condition. Post the waveguide fabrication, we deposited 1 $\mu m$ cladding of $SiO_2$ by plasma etched chemical vapor deposition (PECVD), on top of which, we opened 1-$\mu m$-deep $SiO_2$ window for LC and Al electrodes. 1-$\mu m$-thick Al electrodes were then deposited in the $SiO_2$ windows on both sides of the waveguide by electron-beam evaporation and liftoff. The LC is integrated right above the SiN waveguides allowing evanescent light coupling. We further deposited 200-nm-thick gold metal pads connected to the electrodes for applying electrical signals. The detailed steps on the device fabrication are discussed in the *Methods* section. **Error! Reference source not found.**e shows the microscopic image of different SiN micro ring resonators after LC integration where the separation gap between the metal and waveguide edge is varied. **Error! Reference source not found.**f shows the microscopic image of one of the devices with 1 $\mu$m separation gap between the metal and waveguide edge.

The top view schematic of the LC integrated ring resonator platform with respect to the applied electric field is shown in **Error! Reference source not found.**c and d. As shown in **Error! Reference source not found.**c, the homeotropic alignment layer (see Methods section) aligns the LC molecules vertically to the waveguide with no electric field. The LC molecules can rotate perpendicular to the waveguide by applying an electric field across the electrodes as shown in **Error! Reference source not found.**d. Indeed, as the applied electric field is beyond a certain threshold to overcome the mechanical anchoring of the initial alignment layer, the LC molecules start to rotate to realign to the external electric field. When the field amplitude is large enough, the molecules align themselves perpendicular to the waveguide. On gradually decreasing the electric field across the LC region, the molecules rotate back to their initial anchored orientation. As the LC molecules rotate from an out-of-plane (z-axis) vertical state to an in-plane (x-axis) alignment with the LC director oriented perpendicular to the direction of light propagation in the waveguide, the refractive index of the liquid-crystal medium increases from minimum to maximum, resulting in an increase of the effective index of the waveguide mode, and hence a phase shift[69,95,96].

Since the electric field $E$ is expressed as $E = \frac{V}{d}$, where $V$ is the voltage, and $d$ is the distance

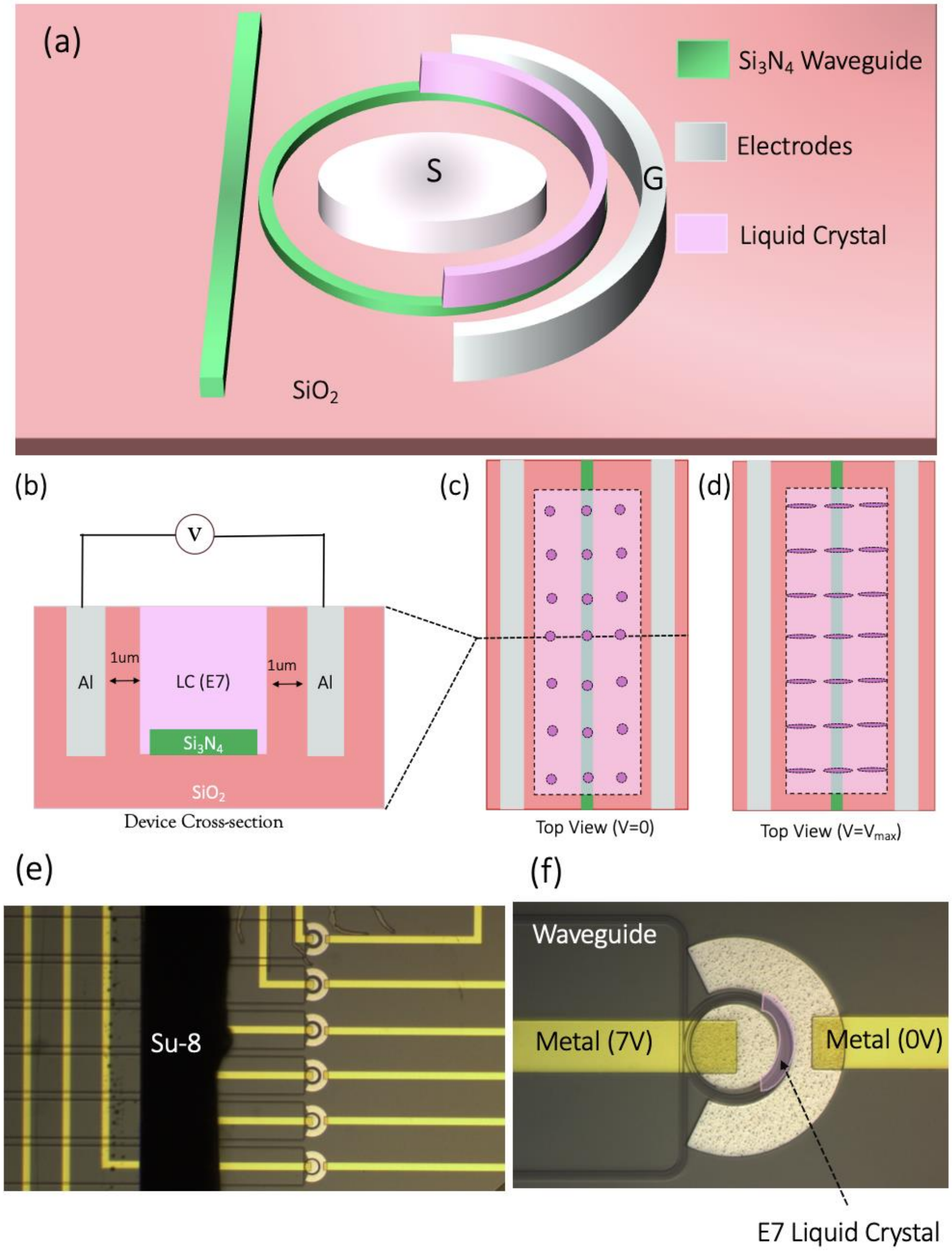

Figure 1: Liquid crystal (LC) integrated silicon nitride micro ring resonator **a.** A schematic of the device **b.** Cross-sectional view of the device **c.** Top view of the device when no external field is applied, and LC molecules are aligned vertically to the waveguide **d.** Top view of the device when LC molecules reorient themselves to align perpendicular to the waveguide subjected to maximum electric field **e.** Microscopic image of devices where the separation gap between the metal and waveguide edge is varied **f.** Microscopic image of the device with 1 $\mu$m metal-waveguide edge separation. The LC region is represented by false color (pink)

between the electrodes, a higher voltage difference across a shorter distance result in a stronger electric field. Therefore, under the same voltage, smaller distance between the metal and the waveguide edge would result in a stronger electric field, and hence a larger phase shift[105,106]. To verify this, we studied the effect of the distance on the induced phase shift both in simulation (Figure 3e) and experiments in Figure 4a, where we indeed observed the expected behavior. We performed all experiments and simulations using transverse electric (TE) mode.

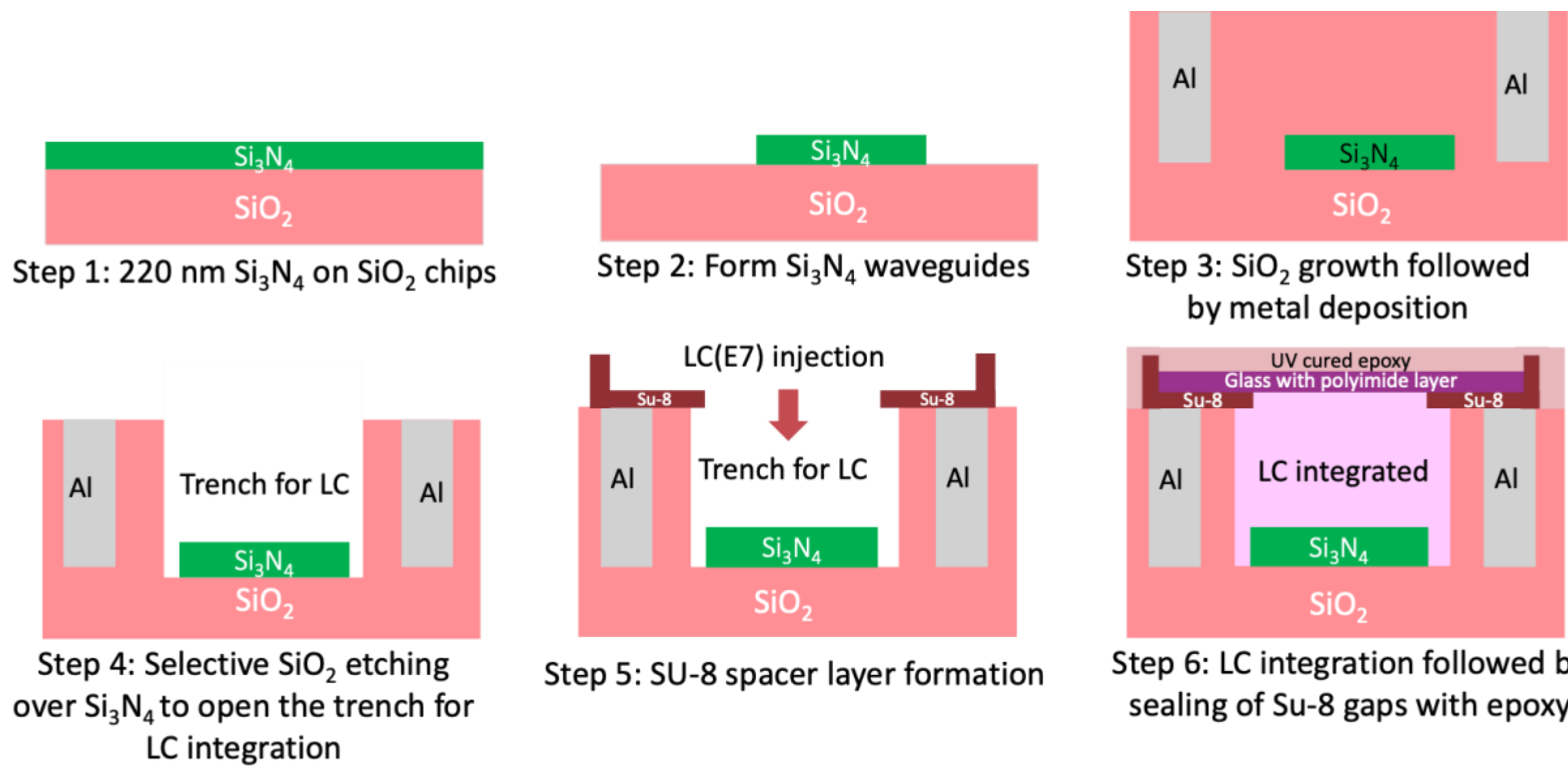

Figure 2: Schematic of the fabrication steps for integrating liquid crystals (LCs) into silicon nitride ring resonator platform at 780 nm

**A. Simulation Studies on LC integrated SiN platform**

We simulated LC integrated SiN waveguides using Finite Difference Eigenmode solver (Lumerical MODE[107,108]) and the results are reported in Figure 3. As demonstrated in Figure 3a and 3b, the LC is right above the SiN waveguide with metals standing alongside the entire LC region to maximize the effective index change ($\Delta n_{eff}$). In the MODE simulation, the distance between the metal and waveguide edge varies from 0.1 $\mu m$ to 1.5 $\mu m$ to study the effects on the effective index, phase shift and loss from E7 LC. The refractive index of the E7 LC is 1.55[109] at ~780 nm when it is aligned vertical to the waveguide (as shown in **Error! Reference source not found.**c) and the effective mode index is ~1.69[109]. By applying an external electric field, the LCs aligns perpendicular to the waveguide (as shown in **Error! Reference source not found.**d) and, due to its birefringence properties, the refractive index increases from 1.55 to 1.69 at 780 nm. From Figure 3b, we observed that in this configuration, the waveguide mode pulled up slightly into the LC, enhancing the interaction between LC and light. Due to the increase of LC refractive index and the mode shape change, we see an increase in effective mode index from 1.70 to 1.73 at 780 nm wavelength.

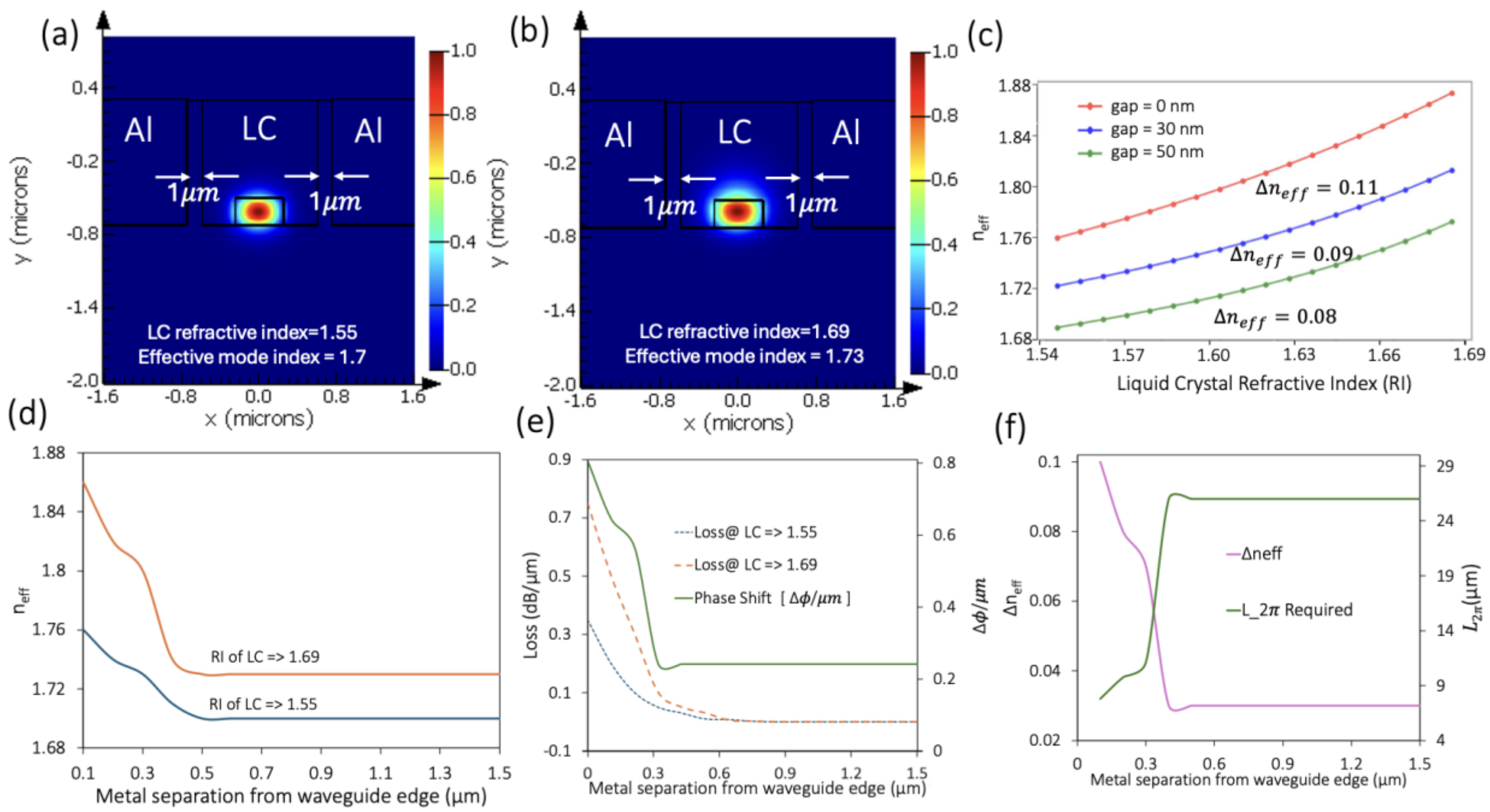

Figure 3: Finite Difference Eigenmode (MODE) Simulation (**a-b**). Optical mode confinement and effective mode index computation at **a.** 1.55 and **b.** 1.69 refractive index of E7 liquid crystal (LC) at 780 nm **c.** Variation of effective mode index ($n_{eff}$) w.r.t. variation of LC refractive index (RI) for varying gap of 0, 30 and 50 nm between the silicon nitride waveguide and LC trench **d.** Variation of $n_{eff}$ w.r.t. variation in metal-waveguide edge separation for LC RI of 1.55 and 1.69 **e.** Variation of loss (dB/$\mu$m) and phase shift per unit length ($\Delta\phi/\mu$m) of modulator w.r.t. variation in metal-waveguide edge separation. $\Delta n_{eff}$: Change in effective mode index **f.** Variation of $\Delta n_{eff}$ and modulator length required to induce a $2\pi$ phase shift w.r.t. variation in metal-waveguide edge separation

We studied the effect of a buffer $SiO_2$ layer between the LC and SiN waveguides in Figure 3c and Supplementary Section 1. Note that here we fixed the distance between metal and waveguide as 0.1 $\mu$m unlike Figure 3a andFigure 3b where the distance between metal and waveguide is fixed at 1 $\mu$m. It shows that the effective index curves shift down and the refractive index contrast ($\Delta n$) decreases as the buffer layer thickness (gap) increases from 0 to 50 nm. Therefore, to maximize the phase shift, we avoid the use of a buffer $SiO_2$ layer between LC and SiN. This can be intuitively explained by the weaker light - LC interaction when a buffer layer exists. In this configuration of 0.1 $\mu$m metal and waveguide separation gap and in absence of buffer $SiO_2$ layer between LC and SiN, the effective mode index increases from 1.76 to 1.87, rendering a change in effective mode index $\Delta n_{eff}$ ~ 0.11 (Figure 3c).

Next, we study the effect of the distance between metal and waveguide in Figure 3d-f. We observed that the effective index contrast decreases with the increase of metal-waveguide distance until around 400 nm, after which the contrast remains constant. This can be explained by the metal helping to extend the optical mode in the *x* direction, which enhances the interaction with the LC. This comes with an unavoidable loss due to metal absorption as shown in Figure 3e, which reduces as the distance increases. When the distance is larger than 500 nm, the excessive loss due to the metal is negligible, and $\Delta n_{eff} \approx 0.03$ between the vertical and perpendicular LC molecules. The change in effective index $\Delta n_{eff}$ can be used to calculate the expected phase shift for a given shifter length using Equation 1[69] Equation 1 .

$$\Delta\phi = \frac{2\pi L \Delta n_{eff}}{\lambda_0} \qquad \text{Equation 1}$$

where, $\Delta\phi$ is the induced phase shift over a phase shifter length L for $\Delta n_{eff}$ change in effective index at free space wavelength $\lambda_0$. Thus, the required phase shifter length to achieve a $2\pi$ phase shift can be calculated using Equation 1, given $\Delta n_{eff} = 0.03$ $\lambda_0 = 780$ nm. The results are plotted in Figure 3f where a phase shifter length of 26 $\mu m$ is required when metallic loss is negligible.

B. **Optical & electrical characterization of fabricated devices**

Although our simulation suggests that no excessive loss due to metal occurs when the metal-waveguide distance is larger than 500 nm, we considered larger metal-waveguide separation gap $\geq 1$ $\mu m$ in the fabricated devices to provide more error margin for lithography overlay. Another possible manufacturing imperfection is related to the removal of $SiO_2$ by fluorine-based dry etching in which the accurate control of the etch depth is difficult in such experiments.

We experimentally validate our simulation results by fabricating LC integrated SiN micro ring resonators with 40-$\mu m$-long LC trench and varying the metal-waveguide distance from 1 to 7 $\mu m$. All micro rings have a bus-ring gap of 100 nm to achieve a near-critical coupling condition. Figure 4a shows the measured optical loss and induced phase shift for varying metal-waveguide distance due to LC. The trend agrees qualitatively well with our simulation, showing a decreasing loss with increased distance. The highest loss and highest phase shift are observed at a metal-waveguide distance of 1 $\mu m$. At 1 $\mu m$ metal-waveguide distance, losses introduced due to the integration of LC into the SiN platform is ~0.08 dB/$\mu m$. The excessive losses due to LC is calculated according to Equation 2[110].

$$\alpha_{0V} = \frac{10}{\log(10)} \cdot \frac{2\pi\lambda_0}{FSR} \cdot \left(\frac{1}{Q_{0V}} - \frac{1}{Q_0}\right) \qquad \text{Equation 2}$$

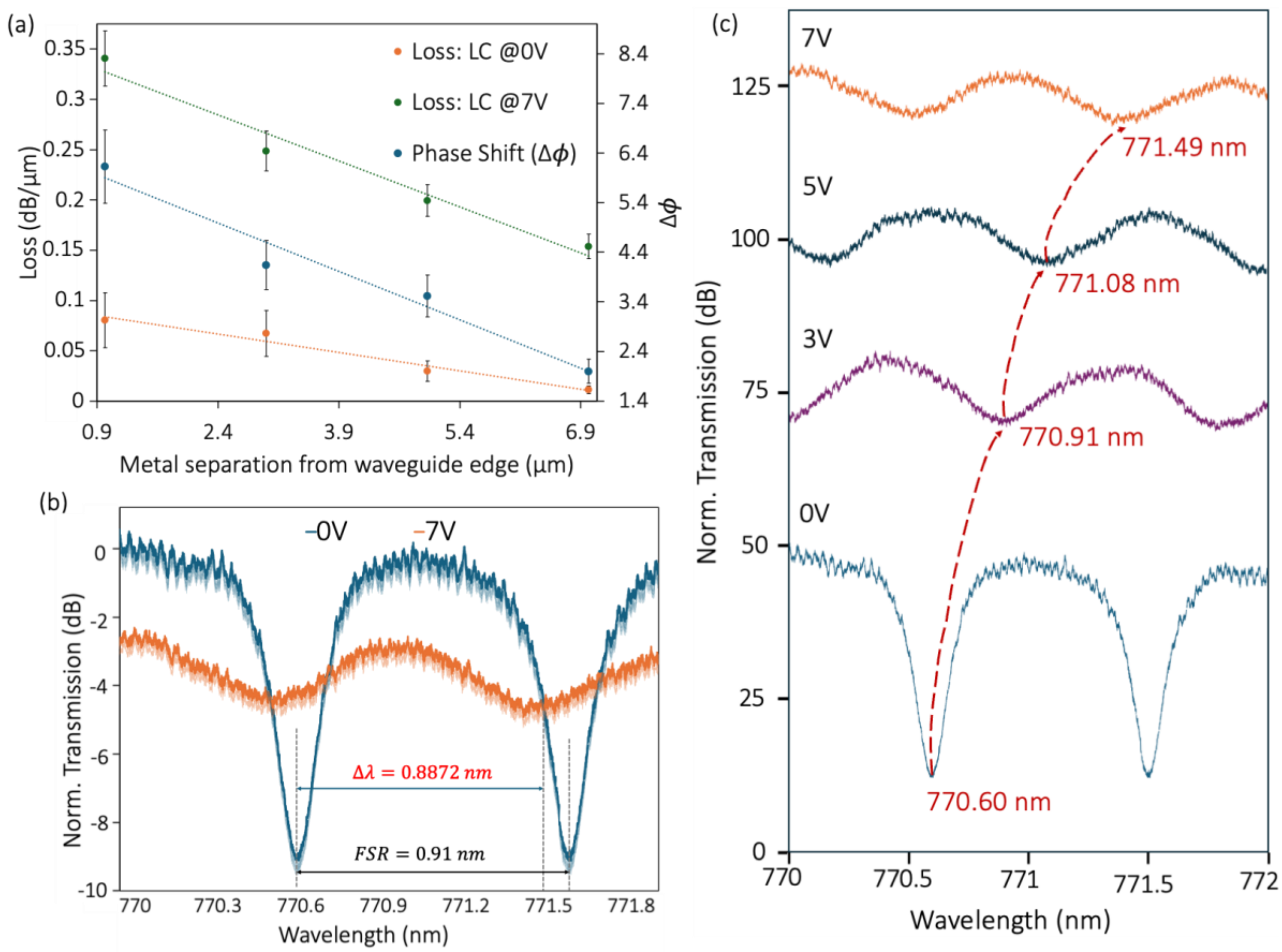

Figure 4: Optical and Electrical Characterization at ~780 nm **a.** Variation of loss and per unit length of modulator (dB/$\mu$m) and total phase shift ($\Delta\phi$) induced along the length of the modulator w.r.t. variation in metal-waveguide edge separation. The additional loss per unit length was calculated after liquid crystal (LC) integration and after LC was subjected to applied electric field (7V). The dashed lines (linear fit to the experimental data) act as a guide to the eye demonstrating the gradual decrease in loss and phase shift with the increase in separation gap between metal and waveguide edge. **b.** Normalized Optical transmission without any external electric field (0V) and after application of external electric field (7V/$\mu$m) across the LC region for metal-waveguide separation gap of 1 $\mu$m. The spectra are averaged over five cycles, and the shadowed area shows the standard deviation. **c.** The observed red shift in resonance from 770.6 nm to 771.49 nm corresponding to ~2$\pi$ phase shift on gradually increasing the voltage from 0V to 7V across the LC region. This shift was also observed in the same device with metal-waveguide separation gap of 1 $\mu$m. Normalized transmission is calculated as Transmission (dB) = 10 ×$\log_{10}$ (Transmitted Power in mW) and the maximum transmission is normalized to 0 dB.

where, $\alpha_{0V}$ is the round-trip optical loss due to the LC integration without external electric field (dB/$\mu m$). $\lambda_0$ ~ 780 nm represents the free space wavelength, FSR represents the free-spectral range which is around 0.91 nm. $Q_0$ and $Q_{0V}$ represent the quality factors before and after the LC integration, respectively. The excessive loss is attributed to both finite absorption loss of LC and scattering losses due to surface roughness at the $Si_3N_4$-LC interfaces.

The additional losses introduced on applying external electric field to the LC is calculated according to Equation 3[110].

$$\alpha_{7V} = \frac{10}{\log(10)} \cdot \frac{2\pi\lambda_0}{FSR} \cdot \left(\frac{1}{Q_{7V}} - \frac{1}{Q_{0V}}\right)$$

Equation 3

where, $\alpha_{7V}$ is the round-trip loss when applying a voltage of ~7 V across the LC. $Q_{7V}$ represents the quality factors with the applied electric field. We extracted an optical loss of ~0.34 dB/$\mu m$ with 7 V, which is attributed to the non-uniform molecular reorientation (due to the circular geometry of the device) and the resulting scattering- and birefringence-related losses[111]. Compared to the 1 $\mu m$ metal-waveguide distance case, when the metal-waveguide distance is 7 $\mu m$, the optical losses are reduced by 75% (from 0.08 to 0.02 dB/$\mu m$) and 47% (from 0.34 to 0.18 dB/$\mu m$) without and with the 7V external voltage, respectively. However, the phase shift also reduces from $\sim 2\pi$ $to$ $\sim 0.64\pi$ when the metal-waveguide distance increases from 1 $\mu m$ to 7 $\mu m$. The optical loss due to LC integration and application of external electric field is higher in experiments than simulations.

We attributed this to fabrication imperfections, such as sidewall roughness in SiN waveguides and non-ideal alignment of LC molecules. However, optimizing the dry etch recipes in SiN waveguide fabrication processes can help achieve smoother waveguide sidewalls that substantially reduce scattering losses[112]. The quality of the $SiO_2$ trench etched for LC integration also plays a significant role; smoother trench surfaces and precise etch-depth control can minimize scattering at the SiN–LC interface[113]. Exploring low loss nematic LC materials[114] and improved alignment coatings can further reduce propagation loss[115]. Recent studies have shown that doping nematic LCs with additives—such as helical carbon nanotubes—can reduce free ion concentration by ~70% and lower dielectric absorption, thereby decreasing optical loss without sacrificing birefringence properties[114]. Moreover, use of photoalignment polymers such as azo-dye coatings can induce uniform LC alignment without mechanical rubbing, thereby eliminating surface-induced scattering loss[115].

The experimental phase shift induced by LCs is estimated using Equation 4[116] where $\Delta\lambda$ represents the resonance shift, FSR is the free spectral range which is 0.91 nm in our devices. As shown in Figure 4a, with the increase in metal-waveguide distance, the phase shift reduces, which matches qualitatively with the simulation. A maximum phase shift of $2\pi$ is observed at a metal-waveguide distance of 1 $\mu m$, which we will focus our characterization and discussion on in later sections.

$$\Delta\phi = \frac{\Delta\lambda}{\mathrm{FSR}} \cdot 2\pi \quad \text{Equation 4}$$

Figure 4(b-c) show the transmission spectra of the 1$\mu m$-distance micro ring resonator with various external electric fields. Upon applying a voltage of 7V, a resonance shift of ~0.8872 nm is obtained (Figure 4b), which is close to a full-FSR (0.91 nm) and indicates a near-$2\pi$ round-trip phase shift. We highlight that the current was only ~0.3 nA, which implies an ultra-low tuning power of ~ 2.1 nW. We repeated ~$2\pi$ optical phase shift multiple times and verified its reversibility and repeatability. Figure 4b presents the spectral variation observed across five cycles, with the shaded region indicating the standard deviation, highlighting excellent repeatability for the binary representation. The voltage required to produce a $\pi$ phase shift ($V_\pi$) along the LC length ($L$) of 40

$\mu m$ is 3.5 V and their product ($V_\pi$ . L) is given by 0.014 V·cm, which significantly outperforms the current state-of-the-art devices, such as thermo-optic, electro-optic phase shifters or piezo-electric phase shifters, typically in the order of a few V·cm as shown in Table 1. Different tuning mechanisms have shown promise for phase control in SiN waveguides—offering fast response and, in some cases, high linearity. However, thermo-optic tuning is power-intensive (~mW) and volatile, electro-optic materials often suffer from low modulation efficiency on SiN, and piezoelectric tuning, while fast, typically requires suspended structures, higher voltages, and consumes more power (~μW–mW). In contrast, our LC-based approach achieves ultralow power consumption (~2.1 nW), exceptional modulation efficiency ($V\pi \cdot L = 0.014$ V·cm), and quasi-static non-volatility, all within a planar, CMOS-compatible process—making it highly suitable for scalable, energy-efficient applications such as reconfigurable photonics, optical memory, and low-refresh-rate beam steering.

The device can also access intermediate levels with different voltages. Figure 4c shows the spectra at various voltages from 0 to 7 V. We can see a gradual shift in resonance towards the right as the voltage gradually increases. This optical and electrical characterization proves the fact that on applying voltage externally, the LC molecules start to rotate and finally align about perpendicularly to the waveguide at a maximum voltage of 7V. It is important to note that increasing external voltage beyond 7V does not further shift the resonance to the right and therefore implies that the re-orientation of the E7 molecules is complete. On gradually reducing the voltage the resonance shifts towards the left and finally returns to ~770.6 nm as the voltage becomes 0V.

Table 1: The $V_\pi . L$ values for current state-of-the-art thermo-optic, electro-optic or piezoelectric phase shifters

| Sl. No. | $V_\pi . L$ (V.cm) | State of the art modulators | Reference |
|---|---|---|---|
| 1 | 43 | Piezoelectric lead zirconate titanate (PZT) on SiN | [117] |
| 2 | 12 | 2D material enabled mid IR optoelectronics | [118] |
| 3 | 10 | GaAs/AlGaAs electrooptic modulator | [82] |
| 4 | 3.2 | PZT films on SiN platform | [119] |
| 5 | 3.2 | Pockels modulators on a silicon nitride platform | [83] |
| 6 | 2.65 | Optical modulators based on thin-film lithium niobate | [120] |
| 7 | 2.2 | LN electro-optic modulators at CMOS voltages | [85] |
| 8 | 1.8 | Nanophotonic lithium niobate electro-optic modulators | [86] |
| 9 | 1.24 | GaGeTe Electro-Optic Device Integrated in Si Photonics | [121] |
| 10 | 1.7 | PZT ring modulator on $SiO_2$/Si substrate | [122] |
| 11 | 1.1 | Lithium niobate (LN) modulator for visible light | [81] |
| 12 | 0.7 | PZT on SiN platform | [123] |
| 13 | 0.55 | Sub-1 Volt electro-optic modulators | [89] |
| 14 | 0.52 | Electro-optic polymer modulator for visible photonics | [88] |
| 15 | 0.45 | Large Pockels effect: barium titanate integrated silicon | [87] |
| 16 | 0.23 | Tellurene: for Mid-infrared Optoelectronics | [124] |
| 17 | 0.2 | $BaTiO_3$ based electro-optic modulator | [125] |
| 18 | 0.052 | ITO-based Mach-Zehnder modulator in Si Photonics | [84] |
| *19* | *0.014* | *Near-visible low power nematic-liquid crystal integrated silicon nitride ring resonator* | *This work* |

### C. Device speed Characterization

We carried out transient behavior measurements on our device to obtain the modulation speed and the results are reported in Figure 5. For speed measurement, we parked the laser at optical resonance wavelength at 0V. Figure 5a shows our measured temporal trace data (blue) of the micro ring resonator in response to a continuous square wave with 50% duty cycle (red) for 100 seconds at a pulsing rate of 100 mHz. The square wave signal has a minimum (maximum) voltage of 0 (7) V to turn on (off) the LC micro ring resonator. As shown in Figure 5a, upon application of the 7V voltage, there is an instantaneous sharp increase in the transmitted signal. This is attributed to the fast dielectric reorientation of the LC molecules to the applied electric field. On applying high voltage, the electric field aligns the LC molecules along its direction, which increases the effective mode index red-shifting the resonance and inducing a $2\pi$ phase shift along the length of the LC region. However, eventually the signal strength reduces and finally saturates representing a 'high' level (binary level '1') or 'on' state. This gradual decay and saturation can be attributed to viscoelastic relaxation, which occurs over milliseconds to seconds, depending on the LC properties and cell configuration. Ionic effects or charge accumulation in the LC layer can introduce additional slow drifts in transmission[126–129].

On removal of the external electric field, the falling time is much longer than the rising time, thus the system takes much longer to reach the 'low' level (binary level '0') or 'off state. Unlike the fast electric-field-induced reorientation, the relaxation of LC molecules takes much longer time to return to their lowest energy state through natural relaxation. The slower decay in transmitted signal can be attributed to weak elastic forces and resistance from LC molecule viscosity. Additionally, effects like surface anchoring, geometry, charge trapping, and thermal fluctuations can further delay full relaxation of the LC molecules[130–132]. Once the LC molecules have fully relaxed, the system stabilizes at the initial resonance condition, where the transmission remains at its lowest level, representing the 'off' state. Interestingly, we observe an initial increase in transmitted signal instantaneously on removing the external voltage (Figure 4a) and then the signal reduces to the 'off' state. The instantaneous increase in transmitted signal upon removing the external voltage can be attributed to transient displacement current. When the voltage is suddenly turned off, the reorientation of the LC molecules induces a transient change in the electric field, leading to a displacement current in the system. This transient current momentarily affects the dielectric relaxation of the LC molecules, causing a temporary increase in effective mode index that briefly increases transmission[133–135]. As the LC gradually return to their low-energy state, the effective mode index decreases, shifting the micro ring resonance back and reducing transmission to the 'off' state. However, it is important to note that while the initial increase in transmission is a transient optical effect, the eventual decay to the 'off' state ensures stable binary switching.

The experiment to characterize the rising time ($t_r$) and falling time ($t_f$) was performed at a lower pulsing rate of 20 mHz to allow enough time to stabilize the response. A PicoScope (details in 'Methods' section) was used to collect the temporal trace data, as shown in Figure 5(b-c). As explained earlier, the rising time '$t_r$' of the device is much faster (~1 ms) compared to the falling time '$t_f$' (~4s). We note that the slow relaxation of LC is not a disadvantage, instead an energy-efficient solution for applications that require memory effects. The slow relaxation of LC ensures that the last-written optical state persists for ~4s even after the voltage is removed, acting as a quasi-static memory[90]. We have observed that the "off" state (after the voltage is removed and the LC has relaxed) remains stable for several hours without any noticeable drift in the optical resonance, based on observations during our experimental runs.

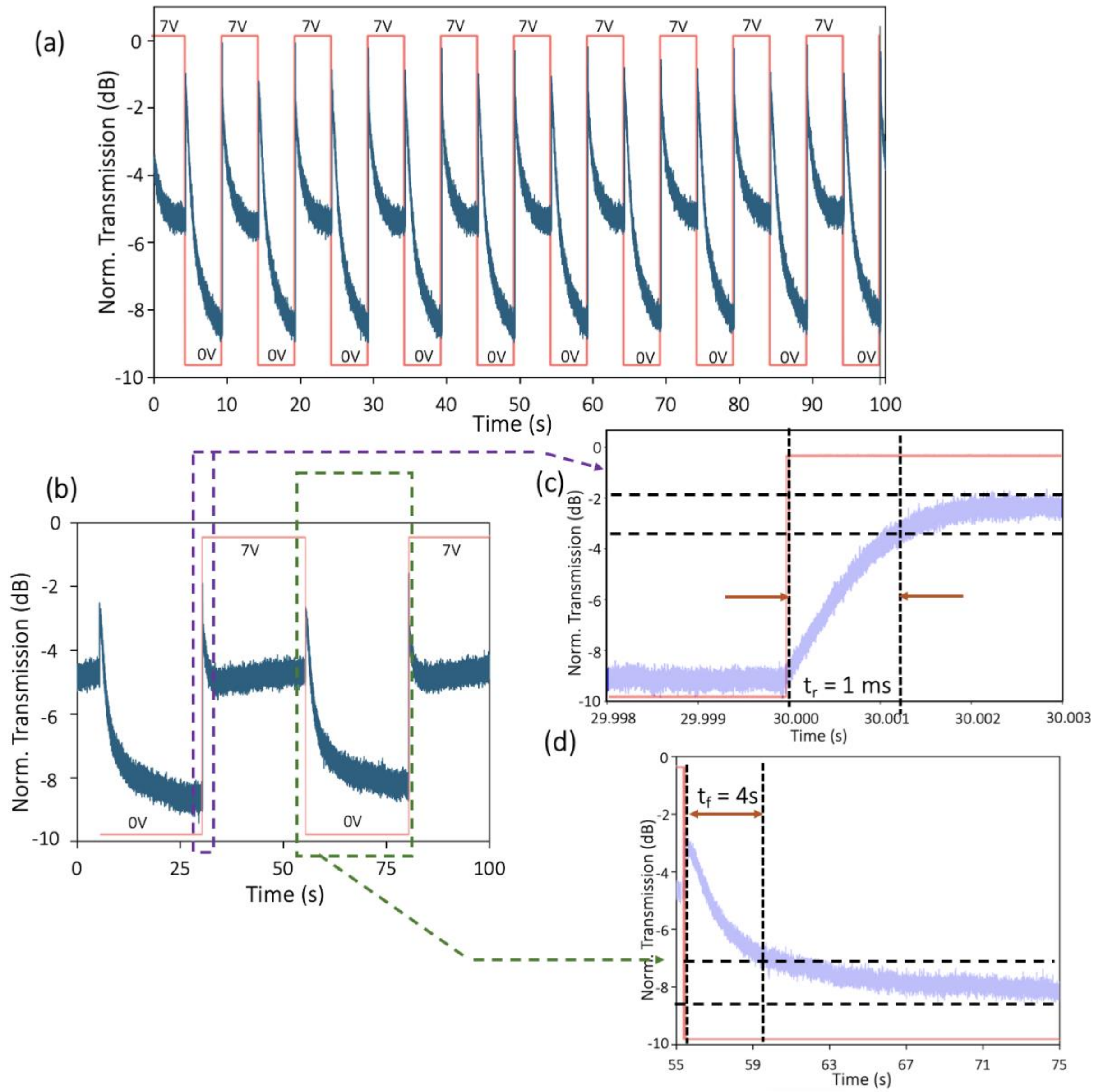

Figure 5: Device response time **a.** The normalized transmission spectra (blue) of the programmable ring resonator in response to a continuous square wave input signal with 50% duty cycle (red) was recorded for 100 seconds at a pulsing rate of 100 mHz. **b.** Normalized transmission observed at a lower pulsing rate of 20 mHz for 100 seconds for 2 cycles such that each cycle of 'high' and 'low' state corresponds to 50 secs. The zoomed in view of the **c.** rise time and **d.** fall time of the device. Norm. Trans. stands for normalized transmission, which is calculated as Transmission (dB) = 10 $\times \log_{10}$ (Transmitted Power in mW) and the maximum transmission is normalized to 0 dB.

This is inherent to the nature of nematic liquid crystals and their relaxation dynamics when not under an applied field and aligned homeotropically, which serves as a stable, lowest-energy state. The slow relaxation (4 seconds for fall time) observed in our device is a key indicator of this quasi-static memory effect. This non-volatile behavior reduces power consumption, as the system does not require continuous voltage or the usage of thin film transistors to maintain the optical state for these 4 seconds. This slow decay ensures stable phase retention, which is crucial for applications like beam steering, holography, and optical computing, where abrupt changes could introduce distortions, side lobes in the steered beam or loss of information. Additionally, the gradual relaxation mitigates flicker and hysteresis effects, ensuring smoother transitions and improved reliability.

The temporal response of our LC-based phase shifter, characterized by a relatively fast rise time of ~1 ms and a much slower fall time of ~4 s, presents a unique profile when compared to other integrated photonic modulator technologies. This asymmetric response is a known characteristic of nematic LC dynamics, where reorientation by an electric field is typically faster than the natural viscoelastic relaxation back to the initial state. While faster fall times can be achieved with thinner LC layers or optimized alignment layers, our chosen geometry and materials provide the memory effect. This non-volatile behavior is a key advantage for applications where the optical state needs to be maintained without continuous power, such as in optical memory or reconfigurable networks that don't require high-frequency symmetrical switching.

*To contextualize our results, we compare the temporal response of our device with recent state-of-the-art LC-based phase shifters integrated into photonic platforms.* Table 2 *summarizes key performance metrics from recent literature, highlighting the diversity in speed, material systems, and operating wavelengths. While some phase shifters demonstrate fast response times via the Pockels effect, they lack non-volatility and require continuous bias (*Table 2*). Other devices based on metasurfaces or slot waveguides achieve moderate response speeds but do not retain phase states without constant actuation (*Table 2*). In contrast, our device combines millisecond-level rise time with long-lived phase retention,*

*making it ideal for applications where energy efficiency and stable phase tuning outweigh ultra-fast modulation requirements. Further, compared to purely thermo-optic*

modulators, our device offers significantly lower power consumption and a much smaller footprint per phase shift due to the large refractive index change in LCs and the ultralow Vπ·L. While electro-optic modulators offer ~nanosecond switching, their higher Vπ·L and integration complexities often make them less suitable for large-scale, low-power arrays, especially in the near-visible spectrum. Thus, our work provides a unique balance, demonstrating a low-power, near-visible LC-integrated SiN phase shifter with excellent Vπ·L and a non-volatile memory feature, making it a promising candidate for quantum and LiDAR applications that can leverage this unique temporal response characteristics.

### D. System-Level Implications and Phase Error Mitigation

The compact size, ultralow power consumption (2.1 nW), and exceptionally low Vπ·L (0.014 V cm) of our liquid crystal integrated SiN phase shifter is highly advantageous for scaling in emerging photonic applications such as quantum optical systems and LiDAR. In quantum optical systems, precise and stable phase control is paramount for maintaining quantum coherence and achieving high-fidelity operations like qubit manipulation and entanglement generation[6,136]. Any phase errors, whether from voltage fluctuations, fabrication imperfections, or minor thermal variations, can degrade these critical functionalities, leading to reduced visibility in interferometric setups or lower quantum gate fidelity[137]. Our low-power approach inherently minimizes thermal effects, a common source of phase instability in other modulator types[39]. Furthermore, the observed slow relaxation behavior (~4s) of the liquid crystal, which results in the last-written optical state persisting even after the voltage is removed, enables a quasi-static memory effect. This non-volatile property is particularly beneficial for quantum applications requiring 'set-and-forget' phase configurations or for energy-efficient reconfigurable networks, as it reduces the need for continuous power to maintain the optical state[63].

For LiDAR systems, which rely on optical phased arrays (OPAs) for non-mechanical beam steering, phase accuracy is directly linked to beam steering precision and side-lobe suppression[5]. Phase errors can lead to distorted beam profiles, reduced target detection accuracy, and increased false detections[17,67]. While the device exhibits a relatively fast turn-on (~1 ms), its slower relaxation time (~4 s) offers a stable, non-volatile state. This characteristic is advantageous for applications requiring a static beam direction for extended periods or for systems that can tolerate slower frame rates in exchange for reduced power consumption and simpler control circuitry, reducing flicker and eliminating the need for continuous refresh cycles[56]. The ultralow power consumption of 2.1 nW per phase shifter is a critical enabler for scaling up to large-element OPAs, where cumulative power budget and thermal management are significant design challenges[25,26].

Table 2: Temporal response comparison of recent LC-based photonic phase shifters

| Sl. No | Platform/Integration | $\lambda$(nm) | LC Material | Rise Time | Fall Time | Key Features | Reference |
|---|---|---|---|---|---|---|---|
| 1 | Inkjet-Printed LC on Si | 1550 | E7 | ~0.6 ms | ~1.2 ms | Scalable; volatile | [138] |
| 2 | LC integrated metasurface | 633 | E7 | ~0.7 ms | ~3.5 ms | 2π phase shift; volatile | [139] |
| 3 | Si slot waveguide | 1550 | 5CB | ~5 ms | ~2 s | Strong modulation, volatile, | [140] |
| 4 | Si ring resonator | 1550 | E7 | ~10 ms | ~2 s | High actuation voltage, volatile | [141] |
| *5* | *SiN ring resonator* | *780* | *E7* | *~1 ms* | *~4 s* | *Not suited for high-speed modulation; Low power, quasi non-volatile, compact (0.014 V-cm)* | *This work* |

**Conclusion**

In summary, we have demonstrated a compact, low-power liquid-crystal-based phase shifter integrated with a SiN micro ring resonator at 780 nm. By incorporating E7 nematic LC, known for its high birefringence (~0.2) and excellent electro-optic properties, our device highlights full-FSR phase tuning capability, a low driving voltage of only 7 V, minimal power consumption (~2 nW), and an extremely small $V_{\pi}\cdot L$ value of 0.014 V·cm. We optimized the metal-waveguide distance as 1 μm to balance the tradeoff between phase shift and optical loss. The measured device exhibits an interesting asymmetric response: a relatively fast rise time (~1 ms), and a much slower fall time (~4 s). This slow relaxation of LC molecules allows the device to retain its optical state even after the voltage is removed without the usage of thin film transistors or latching systems, offering an energy-efficient solution optical memory effects. This work shows a crucial step toward the development of energy-efficient, reconfigurable photonic circuits, such as LiDAR, quantum optical systems, and low-power photonic memory applications.

## Methods

### A. Device Fabrication

The fabrication of silicon photonic devices was conducted on a commercial SiN wafer with a 220 nm thick SiN layer deposited by low pressure chemical vapor deposition (LPCVD) atop a 4 μm thick $SiO_2$ layer (Rogue Valley). Electron beam lithography (EBL, JEOL JBX-6300FS) was used to write the device patterns using a positive-tone EBL resist (ZEP-520A), followed by etching using a fluorine-based inductively coupled plasma etcher (ICP, Oxford Plasma Lab 100 ICP-18) with a mix of SF6 and C4F8 gases, achieving an etch rate of ~ 1.9 nm/sec. This was followed by stripping off the resist using di-methylene chloride solution to get SiN waveguide thickness of ~220nm. Post formation of waveguides, 1 $\mu m$ of silicon dioxide was deposited using plasma enhanced chemical vapor deposition (PECVD). Heidelberg (Heidelberg-DWL66) was used to write the metal regions using positive-tone photoresist AZ1512 followed by fluorine based dry etch of ~1 $\mu m$ of silicon dioxide (etch rate: 5nm/s) using ICP. Metal deposition was achieved via electron-beam evaporation (EVAP, CHA SEC-600) and subsequent lift-off of Aluminum (1 $\mu m$). Metal pads and connecting lines between metal pads and metal regions were written as a separate layer using another Heidelberg-DWL66 write with resist AZ1512 followed by deposition of Ti/Au (15nm/200nm) using EVAP and subsequent liftoff. A 1 $\mu m$ trench was etched into the $SiO_2$ cladding using flouring based ICP etcher for LC integration followed by stripping of the resist using acetone. To ensure the etched trench to be ~1 $\mu m$ a separate test chip was used to calculate the exact etch rate of ~4.75nm/s. A thick SU-8 photoresist layer ~10 $\mu m$ was then patterned on top of the photonic chip followed. The Su-8 is patterned as 2 inverted 'C' with a spacing of 5346 $\mu m$. The Su-8 layer acts as a spacer layer between the glass chip and the photonic chip. On both substrates a LC homeotropic alignment layer is deposited by using a solution of octadecyldimethyl(3-trimethoxysilylpropyl) ammonium chloride (DMOAP), isopropyl alcohol and water with the following volume percentages: 0,1%, 89.9%,10%, respectively. The photonic chip and a glass substate are then immersed vertically in the DMOAP solution at speed rate of 0.4mm/sec followed by a baking in oven at T=120°C for 30 minutes. Finally, the glass and the photonic chip is assembled together, by gluing them in specific points, and the E7 LC is injected followed by sealing the Su-8 gaps with UV cured epoxy. The LC mixture E7, composed of 4-cyano-n-pentyl-biphenyl (5CB), 4-cyano-n-heptyl-biphenyl (7CB), 4-cyano-n-octyloxy-biphenyl (80CB), and 4-cyano-n-pentyl-p-terphenyl (5CT) was commercially purchased and directly infiltrated into the photonic structure.

### B. Lumerical MODE Simulation

All the simulation studies mentioned in the paper including computation of effective mode index, phase shift induced along the length of the modulator and optical losses were performed using a commercial simulation software Lumerical MODE[107,108].

### C. Optical Transmission Measurement Setup

The optical measurement setup consists of vertical fiber array setup angled at 25 degrees, tunable diode laser (Newport TLB-6712), a photodiode and a data acquisition system (DAQ) to measure the static optical transmission. Input light was provided by the laser, and the laser was swept in the range of 765 nm to 780 nm. The optical fibers were coupled to the on-chip gratings and maximum fiber-to chip coupling efficiency was ensured from the photodiode response. For on-chip external electric field, electrical pulses were applied to the on-chip metal contacts via a pair of electrical probes positioned with probe positioners (Cascade Microtech DPP105-M-AI-S). DC measurements were performed by applying voltage via Keithley and AC measurements were performed by applying square pulses via function arbitrary generator (Keysight 81160 A). A 200 MHz 4-Channel Mixed-Signal Oscilloscope (PicoScope, Pico 5444D MSO) was used to was used to capture the temporal trace data. The tunable laser and photodiode were controlled through a LabView program. A glimpse of the measurement setup with the fabricated chip is shown in Supplementary Figure 2.

### Supplementary Material

The supplementary material includes additional information on optical and electrical characterization of E7 liquid crystal integrated SiN ring resonator.

### Acknowledgements

The research is funded by the NASA and DHS STTR grants. Additionally, this material is based upon work supported by the Defense Health Agency (DHA) Small Business Innovation Research (SBIR)/ Small Business Technology Transfer (STTR) Programs/U.S. Army Research Institute of Environmental Medicine (USARIEM) under U.S. Army Medical Research Acquisition Activity (USAMRAA) Contract No HT942524C0068. Part of this work was conducted at the Washington Nanofabrication Facility/Molecular Analysis Facility, a National Nanotechnology Coordinated Infrastructure (NNCI) site at the University of Washington. Any opinions, findings and conclusions or recommendations expressed in this material are those of the author(s) and do not necessarily reflect the views of the Department of Defense, DHA, SBIR/STTR Programs, USARIEM, or USAMRAA.

**References**

(1) Christen, I.; Propson, T.; Sutula, M.; Sattari, H.; Choong, G.; Panuski, C.; Melville, A.; Mallek, J.; Brabec, C.; Hamilton, S.; Dixon, P. B.; Menssen, A. J.; Braje, D.; Ghadimi, A. H.; Englund, D. An Integrated Photonic Engine for Programmable Atomic Control. *Nat. Commun.* **2025**, *16* (1), 82. https://doi.org/10.1038/s41467-024-55423-3.

(2) *Integrated photonics beyond communications | Applied Physics Letters | AIP Publishing*. https://pubs.aip.org/aip/apl/article/123/23/230501/2926235/Integrated-photonics-beyond-communications (accessed 2025-03-10).

(3) Butt, M. A.; Janaszek, B.; Piramidowicz, R. Lighting the Way Forward: The Bright Future of Photonic Integrated Circuits. *Sens. Int.* **2025**, *6*, 100326. https://doi.org/10.1016/j.sintl.2025.100326.

(4) Shekhar, S.; Bogaerts, W.; Chrostowski, L.; Bowers, J. E.; Hochberg, M.; Soref, R.; Shastri, B. J. Roadmapping the next Generation of Silicon Photonics. *Nat. Commun.* **2024**, *15* (1), 751. https://doi.org/10.1038/s41467-024-44750-0.

(5) Dostart, N.; Zhang, B.; Khilo, A.; Brand, M.; Qubaisi, K. A.; Onural, D.; Feldkhun, D.; Wagner, K. H.; Popović, M. A. Serpentine Optical Phased Arrays for Scalable Integrated Photonic Lidar Beam Steering. *Optica* **2020**, *7* (6), 726–733. https://doi.org/10.1364/OPTICA.389006.

(6) Feng, L.; Zhang, M.; Wang, J.; Zhou, X.; Qiang, X.; Guo, G.; Ren, X. Silicon Photonic Devices for Scalable Quantum Information Applications. *Photonics Res.* **2022**, *10* (10), A135–A153. https://doi.org/10.1364/PRJ.464808.

(7) Mahmudlu, H.; Johanning, R.; van Rees, A.; Khodadad Kashi, A.; Epping, J. P.; Haldar, R.; Boller, K.-J.; Kues, M. Fully On-Chip Photonic Turnkey Quantum Source for Entangled Qubit/Qudit State Generation. *Nat. Photonics* **2023**, *17* (6), 518–524. https://doi.org/10.1038/s41566-023-01193-1.

(8) *Chip-scale nonlinear photonics for quantum light generation | AVS Quantum Science | AIP Publishing*. https://pubs.aip.org/avs/aqs/article/2/4/041702/997293/Chip-scale-nonlinear-photonics-for-quantum-light (accessed 2025-03-10).

(9) Lukens, J. M.; Lougovski, P. Frequency-Encoded Photonic Qubits for Scalable Quantum Information Processing. *Optica* **2017**, *4* (1), 8–16. https://doi.org/10.1364/OPTICA.4.000008.
(10) Jia, X.; Zhai, C.; Zhu, X.; You, C.; Cao, Y.; Zhang, X.; Zheng, Y.; Fu, Z.; Mao, J.; Dai, T.; Chang, L.; Su, X.; Gong, Q.; Wang, J. Continuous-Variable Multipartite Entanglement in an Integrated Microcomb. *Nature* **2025**, 1–8. https://doi.org/10.1038/s41586-025-08602-1.
(11) Kuzmenko, K.; Vines, P.; Halimi, A.; Collins, R. J.; Maccarone, A.; McCarthy, A.; Greener, Z. M.; Kirdoda, J.; Dumas, D. C. S.; Llin, L. F.; Mirza, M. M.; Millar, R. W.; Paul, D. J.; Buller, G. S. 3D LIDAR Imaging Using Ge-on-Si Single–Photon Avalanche Diode Detectors. *Opt. Express* **2020**, *28* (2), 1330–1344. https://doi.org/10.1364/OE.383243.
(12) Morimoto, K.; Ardelean, A.; Wu, M.-L.; Ulku, A. C.; Antolovic, I. M.; Bruschini, C.; Charbon, E. Megapixel Time-Gated SPAD Image Sensor for 2D and 3D Imaging Applications. *Optica* **2020**, *7* (4), 346–354. https://doi.org/10.1364/OPTICA.386574.
(13) Shin, D.; Xu, F.; Venkatraman, D.; Lussana, R.; Villa, F.; Zappa, F.; Goyal, V. K.; Wong, F. N. C.; Shapiro, J. H. Photon-Efficient Imaging with a Single-Photon Camera. *Nat. Commun.* **2016**, *7* (1), 12046. https://doi.org/10.1038/ncomms12046.
(14) Behroozpour, B.; Sandborn, P. A. M.; Wu, M. C.; Boser, B. E. Lidar System Architectures and Circuits. *IEEE Commun. Mag.* **2017**, *55* (10), 135–142. https://doi.org/10.1109/MCOM.2017.1700030.
(15) Javidi, B.; Carnicer, A.; Arai, J.; Fujii, T.; Hua, H.; Liao, H.; Martínez-Corral, M.; Pla, F.; Stern, A.; Waller, L.; Wang, Q.-H.; Wetzstein, G.; Yamaguchi, M.; Yamamoto, H. Roadmap on 3D Integral Imaging: Sensing, Processing, and Display. *Opt. Express* **2020**, *28* (22), 32266–32293. https://doi.org/10.1364/OE.402193.
(16) *Photonic technologies for autonomous cars: feature introduction*. https://opg.optica.org/oe/fulltext.cfm?uri=oe-27-5-7627&id=406852 (accessed 2025-03-10).
(17) Rogers, C.; Piggott, A. Y.; Thomson, D. J.; Wiser, R. F.; Opris, I. E.; Fortune, S. A.; Compston, A. J.; Gondarenko, A.; Meng, F.; Chen, X.; Reed, G. T.; Nicolaescu, R. A Universal 3D Imaging Sensor on a Silicon Photonics Platform. *Nature* **2021**, *590* (7845), 256–261. https://doi.org/10.1038/s41586-021-03259-y.
(18) Schwarz, B. Mapping the World in 3D. *Nat. Photonics* **2010**, *4* (7), 429–430. https://doi.org/10.1038/nphoton.2010.148.
(19) Juliano Martins, R.; Marinov, E.; Youssef, M. A. B.; Kyrou, C.; Joubert, M.; Colmagro, C.; Gâté, V.; Turbil, C.; Coulon, P.-M.; Turover, D.; Khadir, S.; Giudici, M.; Klitis, C.; Sorel, M.; Genevet, P. Metasurface-Enhanced Light Detection and Ranging Technology. *Nat. Commun.* **2022**, *13* (1), 5724. https://doi.org/10.1038/s41467-022-33450-2.
(20) Yi, Y.; Wu, D.; Kakdarvishi, V.; Yu, B.; Zhuang, Y.; Khalilian, A. Photonic Integrated Circuits for an Optical Phased Array. *Photonics* **2024**, *11* (3), 243. https://doi.org/10.3390/photonics11030243.
(21) Doylend, J. K.; Heck, M. J. R.; Bovington, J. T.; Peters, J. D.; Coldren, L. A.; Bowers, J. E. Two-Dimensional Free-Space Beam Steering with an Optical Phased Array on Silicon-on-Insulator. *Opt. Express* **2011**, *19* (22), 21595–21604. https://doi.org/10.1364/OE.19.021595.

(22) Guo, W.; Binetti, P. R. A.; Althouse, C.; Mašanović, M. L.; Ambrosius, H. P. M. M.; Johansson, L. A.; Coldren, L. A. Two-Dimensional Optical Beam Steering With InP-Based Photonic Integrated Circuits. *IEEE J. Sel. Top. Quantum Electron.* **2013**, *19* (4), 6100212–6100212. https://doi.org/10.1109/JSTQE.2013.2238218.
(23) Guo, Y.; Guo, Y.; Li, C.; Zhang, H.; Zhou, X.; Zhang, L. Integrated Optical Phased Arrays for Beam Forming and Steering. *Appl. Sci.* **2021**, *11* (9), 4017. https://doi.org/10.3390/app11094017.
(24) Aflatouni, F.; Abiri, B.; Rekhi, A.; Hajimiri, A. Nanophotonic Projection System. *Opt. Express* **2015**, *23* (16), 21012–21022. https://doi.org/10.1364/OE.23.021012.
(25) Sun, J.; Timurdogan, E.; Yaacobi, A.; Hosseini, E. S.; Watts, M. R. Large-Scale Nanophotonic Phased Array. *Nature* **2013**, *493* (7431), 195–199. https://doi.org/10.1038/nature11727.
(26) Tossoun, B.; Xiao, X.; Cheung, S.; Yuan, Y.; Peng, Y.; Srinivasan, S.; Giamougiannis, G.; Huang, Z.; Singaraju, P.; London, Y.; Hejda, M.; Sundararajan, S. P.; Hu, Y.; Gong, Z.; Baek, J.; Descos, A.; Kapusta, M.; Böhm, F.; Van Vaerenbergh, T.; Fiorentino, M.; Kurczveil, G.; Liang, D.; Beausoleil, R. G. Large-Scale Integrated Photonic Device Platform for Energy-Efficient AI/ML Accelerators. *IEEE J. Sel. Top. Quantum Electron.* **2025**, *31* (3: AI/ML Integrated Opto-electronics), 1–26. https://doi.org/10.1109/JSTQE.2025.3527904.
(27) Xu, D.; Ma, Y.; Jin, G.; Cao, L. Intelligent Photonics: A Disruptive Technology to Shape the Present and Redefine the Future. *Engineering* **2024**. https://doi.org/10.1016/j.eng.2024.08.016.
(28) *Photonics for Neuromorphic Computing: Fundamentals, Devices, and Opportunities - Li - 2025 - Advanced Materials - Wiley Online Library*. https://advanced.onlinelibrary.wiley.com/doi/full/10.1002/adma.202312825 (accessed 2025-03-10).
(29) Pérez-López, D.; Gutierrez, A.; Sánchez, D.; López-Hernández, A.; Gutierrez, M.; Sánchez-Gomáriz, E.; Fernández, J.; Cruz, A.; Quirós, A.; Xie, Z.; Benitez, J.; Bekesi, N.; Santomé, A.; Pérez-Galacho, D.; DasMahapatra, P.; Macho, A.; Capmany, J. General-Purpose Programmable Photonic Processor for Advanced Radiofrequency Applications. *Nat. Commun.* **2024**, *15* (1), 1563. https://doi.org/10.1038/s41467-024-45888-7.
(30) Shi, Y.; Zhang, Y.; Wan, Y.; Yu, Y.; Zhang, Y.; Hu, X.; Xiao, X.; Xu, H.; Zhang, L.; Pan, B. Silicon Photonics for High-Capacity Data Communications. *Photonics Res.* **2022**, *10* (9), A106–A134. https://doi.org/10.1364/PRJ.456772.
(31) Chen, R.; Tara, V.; Choi, M.; Dutta, J.; Sim, J.; Ye, J.; Fang, Z.; Zheng, J.; Majumdar, A. Deterministic Quasi-Continuous Tuning of Phase-Change Material Integrated on a High-Volume 300-Mm Silicon Photonics Platform. *Npj Nanophotonics* **2024**, *1* (1), 1–9. https://doi.org/10.1038/s44310-024-00009-6.
(32) Chen, R.; Tara, V.; Choi, M.; Duta, J.; Sim, J.; Ye, J.; Zheng, J.; Fang, Z.; Majumdar, A. Toward Very-Large-Scale Nonvolatile Electrically Programmable Photonic Integrated Circuits with Deterministic Multilevel Operation. In *CLEO 2024 (2024), paper AM1J.5*; Optica Publishing Group, 2024; p AM1J.5. https://doi.org/10.1364/CLEO_AT.2024.AM1J.5.

(33) Mitchell, C. J.; Hu, T.; Sun, S.; Stirling, C. J.; Nedeljkovic, M.; Peacock, A. C.; Reed, G. T.; Mashanovich, G. Z.; Rowe, D. J. Mid-Infrared Silicon Photonics: From Benchtop to Real-World Applications. *APL Photonics* **2024**, *9* (8), 080901. https://doi.org/10.1063/5.0222890.
(34) Quack, N.; Takabayashi, A. Y.; Sattari, H.; Edinger, P.; Jo, G.; Bleiker, S. J.; Errando-Herranz, C.; Gylfason, K. B.; Niklaus, F.; Khan, U.; Verheyen, P.; Mallik, A. K.; Lee, J. S.; Jezzini, M.; Zand, I.; Morrissey, P.; Antony, C.; O'Brien, P.; Bogaerts, W. Integrated Silicon Photonic MEMS. *Microsyst. Nanoeng.* **2023**, *9* (1), 1–22. https://doi.org/10.1038/s41378-023-00498-z.
(35) Poon, J. K. S.; Govdeli, A.; Sharma, A.; Mu, X.; Chen, F.-D.; Xue, T.; Liu, T. Silicon Photonics for the Visible and Near-Infrared Spectrum. *Adv. Opt. Photonics* **2024**, *16* (1), 1–59. https://doi.org/10.1364/AOP.501846.
(36) Chen, R.; Tara, V.; Dutta, J.; Fang, Z.; Zheng, J.; Majumdar, A. Low-Loss Multilevel Operation Using Lossy Phase-Change Material-Integrated Silicon Photonics. *J. Opt. Microsyst.* **2024**, *4* (3), 031202. https://doi.org/10.1117/1.JOM.4.3.031202.
(37) Beliaev, L. Yu.; Shkondin, E.; Lavrinenko, A. V.; Takayama, O. Optical, Structural and Composition Properties of Silicon Nitride Films Deposited by Reactive Radio-Frequency Sputtering, Low Pressure and Plasma-Enhanced Chemical Vapor Deposition. *Thin Solid Films* **2022**, *763*, 139568. https://doi.org/10.1016/j.tsf.2022.139568.
(38) Corato-Zanarella, M.; Ji, X.; Mohanty, A.; Lipson, M. Absorption and Scattering Limits of Silicon Nitride Integrated Photonics in the Visible Spectrum. *Opt. Express* **2024**, *32* (4), 5718–5728. https://doi.org/10.1364/OE.505892.
(39) Ortmann, J. E.; Eltes, F.; Caimi, D.; Meier, N.; Demkov, A. A.; Czornomaz, L.; Fompeyrine, J.; Abel, S. Ultra-Low-Power Tuning in Hybrid Barium Titanate–Silicon Nitride Electro-Optic Devices on Silicon. *ACS Photonics* **2019**, *6* (11), 2677–2684. https://doi.org/10.1021/acsphotonics.9b00558.
(40) Gardes, F.; Shooa, A.; De Paoli, G.; Skandalos, I.; Ilie, S.; Rutirawut, T.; Talataisong, W.; Faneca, J.; Vitali, V.; Hou, Y.; Bucio, T. D.; Zeimpekis, I.; Lacava, C.; Petropoulos, P. A Review of Capabilities and Scope for Hybrid Integration Offered by Silicon-Nitride-Based Photonic Integrated Circuits. *Sensors* **2022**, *22* (11), 4227. https://doi.org/10.3390/s22114227.
(41) Zhang, Y.; Guo, X.; Ji, X.; Shen, J.; He, A.; Su, Y. What Can Be Integrated on the Silicon Photonics Platform and How? *APL Photonics* **2024**, *9* (9), 090902. https://doi.org/10.1063/5.0220463.
(42) Sacher, W. D.; Huang, Y.; Lo, G.-Q.; Poon, J. K. S. Multilayer Silicon Nitride-on-Silicon Integrated Photonic Platforms and Devices. *J. Light. Technol.* **2015**, *33* (4), 901–910.
(43) *A Thermally Tunable 1 × 4 Channel Wavelength Demultiplexer Designed on a Low-Loss Si3N4 Waveguide Platform*. https://www.mdpi.com/2304-6732/2/4/1065 (accessed 2025-03-10).
(44) Shtyrkova, K.; Callahan, P. T.; Li, N.; Magden, E. S.; Ruocco, A.; Vermeulen, D.; Kärtner, F. X.; Watts, M. R.; Ippen, E. P. Integrated CMOS-Compatible Q-Switched Mode-Locked Lasers at 1900nm with an on-Chip Artificial Saturable Absorber. *Opt. Express* **2019**, *27* (3), 3542–3556. https://doi.org/10.1364/OE.27.003542.

(45) Wu, Q.; Zhang, H.; Jia, D.; Liu, T. Recent Development of Tunable Optical Devices Based on Liquid. *Molecules* **2022**, *27* (22), 8025. https://doi.org/10.3390/molecules27228025.
(46) Zhang, Z.; You, Z.; Chu, D. Fundamentals of Phase-Only Liquid Crystal on Silicon (LCOS) Devices. *Light Sci. Appl.* **2014**, *3* (10), e213–e213. https://doi.org/10.1038/lsa.2014.94.
(47) Badloe, T.; Kim, J.; Kim, I.; Kim, W.-S.; Kim, W. S.; Kim, Y.-K.; Rho, J. Liquid Crystal-Powered Mie Resonators for Electrically Tunable Photorealistic Color Gradients and Dark Blacks. *Light Sci. Appl.* **2022**, *11* (1), 118. https://doi.org/10.1038/s41377-022-00806-8.
(48) Notaros, M.; DeSantis, D. M.; Raval, M.; Notaros, J. Liquid-Crystal-Based Visible-Light Integrated Optical Phased Arrays and Application to Underwater Communications. *Opt. Lett.* **2023**, *48* (20), 5269–5272. https://doi.org/10.1364/OL.494387.
(49) Otieno, E.; Matczyszyn, K.; Mokaya, N. Liquid Crystals: Unlocking the Quantum Revolution in Computing. Chemistry August 23, 2024. https://doi.org/10.26434/chemrxiv-2024-ql0fx.
(50) *The diverse world of liquid crystals | Physics Today | AIP Publishing*. https://pubs.aip.org/physicstoday/article/60/9/54/395691/The-diverse-world-of-liquid (accessed 2025-03-10).
(51) *Liquid Crystal Beam Steering Devices: Principles, Recent Advances, and Future Developments*. https://www.mdpi.com/2073-4352/9/6/292 (accessed 2025-03-10).
(52) Mansha, S.; Moitra, P.; Xu, X.; Mass, T. W. W.; Veetil, R. M.; Liang, X.; Li, S.-Q.; Paniagua-Domínguez, R.; Kuznetsov, A. I. High Resolution Multispectral Spatial Light Modulators Based on Tunable Fabry-Perot Nanocavities. *Light Sci. Appl.* **2022**, *11* (1), 141. https://doi.org/10.1038/s41377-022-00832-6.
(53) Morris, R.; Jones, C.; Nagaraj, M. Liquid Crystal Devices for Beam Steering Applications. *Micromachines* **2021**, *12*. https://doi.org/10.3390/mi12030247.
(54) *Beyond the display: phase-only liquid crystal on Silicon devices and their applications in photonics [Invited]*. https://opg.optica.org/oe/fulltext.cfm?uri=oe-27-11-16206&id=412885 (accessed 2025-03-10).
(55) Ma, L.-L.; Li, C.-Y.; Pan, J.-T.; Ji, Y.-E.; Jiang, C.; Zheng, R.; Wang, Z.-Y.; Wang, Y.; Li, B.-X.; Lu, Y.-Q. Self-Assembled Liquid Crystal Architectures for Soft Matter Photonics. *Light Sci. Appl.* **2022**, *11* (1), 270. https://doi.org/10.1038/s41377-022-00930-5.
(56) Beddoe, M.; Walden, S. L.; Miljevic, S.; Pidgayko, D.; Zou, C.; Minovich, A. E.; Barreda, A.; Pertsch, T.; Staude, I. Spatially Controlled All-Optical Switching of Liquid-Crystal-Empowered Metasurfaces. *ACS Photonics* **2025**, *12* (2), 963–970. https://doi.org/10.1021/acsphotonics.4c02029.
(57) Slussarenko, S.; Piccirillo, B.; Chigrinov, V.; Marrucci, L.; Santamato, E. Liquid Crystal Spatial-Mode Converters for the Orbital Angular Momentum of Light. *J. Opt.* **2013**, *15* (2), 025406. https://doi.org/10.1088/2040-8978/15/2/025406.
(58) *Four-bit input linear optical quantum computing with liquid crystal devices | APL Quantum | AIP Publishing*.

https://pubs.aip.org/aip/apq/article/1/4/046105/3317725/Four-bit-input-linear-optical-quantum-computing (accessed 2025-03-10).
(59) *Quantum dots for photonic quantum information technology*. https://opg.optica.org/aop/fulltext.cfm?uri=aop-15-3-613&id=537017 (accessed 2025-03-10).
(60) Sultanov, V.; Kavčič, A.; Kokkinakis, E.; Sebastián, N.; Chekhova, M. V.; Humar, M. Tunable Entangled Photon-Pair Generation in a Liquid Crystal. *Nature* **2024**, *631* (8020), 294–299. https://doi.org/10.1038/s41586-024-07543-5.
(61) de Blas, M. G.; García, J. P.; Andreu, S. V.; Arregui, X. Q.; Caño-García, M.; Geday, M. A. High Resolution 2D Beam Steerer Made from Cascaded 1D Liquid Crystal Phase Gratings. *Sci. Rep.* **2022**, *12* (1), 5145. https://doi.org/10.1038/s41598-022-09201-0.
(62) Lindle, J. R.; Watnik, A. T.; Cassella, V. A. Efficient Multibeam Large-Angle Nonmechanical Laser Beam Steering from Computer-Generated Holograms Rendered on a Liquid Crystal Spatial Light Modulator. *Appl. Opt.* **2016**, *55* (16), 4336–4341. https://doi.org/10.1364/AO.55.004336.
(63) Kim, J.; Oh, C.; Serati, S.; Escuti, M. J. Wide-Angle, Nonmechanical Beam Steering with High Throughput Utilizing Polarization Gratings. *Appl. Opt.* **2011**, *50* (17), 2636–2639. https://doi.org/10.1364/AO.50.002636.
(64) Resler, D. P.; Hobbs, D. S.; Sharp, R. C.; Friedman, L. J.; Dorschner, T. A. High-Efficiency Liquid-Crystal Optical Phased-Array Beam Steering. *Opt. Lett.* **1996**, *21* (9), 689–691. https://doi.org/10.1364/OL.21.000689.
(65) *Dynamic Beam Switching by Liquid Crystal Tunable Dielectric Metasurfaces | ACS Photonics*. https://pubs.acs.org/doi/10.1021/acsphotonics.7b01343 (accessed 2025-03-10).
(66) Li, S.-Q.; Xu, X.; Maruthiyodan Veetil, R.; Valuckas, V.; Paniagua-Domínguez, R.; Kuznetsov, A. I. Phase-Only Transmissive Spatial Light Modulator Based on Tunable Dielectric Metasurface. *Science* **2019**, *364* (6445), 1087–1090. https://doi.org/10.1126/science.aaw6747.
(67) Mcmanamon, P.; Bos, P.; Escuti, M.; Heikenfeld, J.; Serati, S.; Xie, H.; Watson, E. A Review of Phased Array Steering for Narrow-Band Electrooptical Systems. *Proc. IEEE* **2009**, *97*, 1078–1096. https://doi.org/10.1109/JPROC.2009.2017218.
(68) Mur, U.; Ravnik, M.; Seč, D. Controllable Shifting, Steering, and Expanding of Light Beam Based on Multi-Layer Liquid-Crystal Cells. *Sci. Rep.* **2022**, *12* (1), 352. https://doi.org/10.1038/s41598-021-04164-0.
(69) Notaros, M.; Dyer, T.; Raval, M.; Baiocco, C.; Notaros, J.; Watts, M. R. Integrated Visible-Light Liquid-Crystal-Based Phase Modulators. *Opt. Express* **2022**, *30* (8), 13790–13801. https://doi.org/10.1364/OE.454494.
(70) Cort, W. D.; Beeckman, J.; Claes, T.; Neyts, K.; Baets, R. Wide Tuning of Silicon-on-Insulator Ring Resonators with a Liquid Crystal Cladding. *Opt. Lett.* **2011**, *36* (19), 3876–3878. https://doi.org/10.1364/OL.36.003876.
(71) Cort, W. D.; Beeckman, J.; James, R.; Fernández, F. A.; Baets, R.; Neyts, K. Tuning of Silicon-on-Insulator Ring Resonators with Liquid Crystal Cladding Using the Longitudinal Field Component. *Opt. Lett.* **2009**, *34* (13), 2054–2056. https://doi.org/10.1364/OL.34.002054.

(72) Ptasinski, J.; Kim, S. W.; Pang, L.; Khoo, I.-C.; Fainman, Y. Optical Tuning of Silicon Photonic Structures with Nematic Liquid Crystal Claddings. *Opt. Lett.* **2013**, *38* (12), 2008–2010. https://doi.org/10.1364/OL.38.002008.
(73) Wang, T.-J.; Li, W.-J.; Chen, T.-J. Radially Realigning Nematic Liquid Crystal for Efficient Tuning of Microring Resonators. *Opt. Express* **2013**, *21* (23), 28974–28979. https://doi.org/10.1364/OE.21.028974.
(74) Dai, J.; Zhang, M.; Zhou, F.; Wang, Y.; Lu, L.; Liu, D. Efficiently Tunable and Fabrication Tolerant Double-Slot Microring Resonators Incorporating Nematic Liquid Crystal as Claddings. *Opt. Commun.* **2015**, *350*, 235–240. https://doi.org/10.1016/j.optcom.2015.02.026.
(75) *Electrically tunable ring resonators incorporating nematic liquid crystals as cladding layers | Applied Physics Letters | AIP Publishing*. https://pubs.aip.org/aip/apl/article/83/23/4689/115808/Electrically-tunable-ring-resonators-incorporating (accessed 2025-03-10).
(76) Falco, A. D.; Assanto, G. Tunable Wavelength-Selective Add–Drop in Liquid Crystals on a Silicon Microresonator. *Opt. Commun.* **2007**, *279* (1), 210–213. https://doi.org/10.1016/j.optcom.2007.06.063.
(77) Chiang, L.-Y.; Wang, C.-T.; Pappert, S.; Yu, P. K. L. Silicon–Organic Hybrid Thermo-Optic Switch Based on a Slot Waveguide Directional Coupler. *Opt. Lett.* **2022**, *47* (15), 3940–3943. https://doi.org/10.1364/OL.467858.
(78) Atsumi, Y.; Watabe, K.; Uda, N.; Miura, N.; Sakakibara, Y. Initial Alignment Control Technique Using On-Chip Groove Arrays for Liquid Crystal Hybrid Silicon Optical Phase Shifters. *Opt. Express* **2019**, *27* (6), 8756–8767. https://doi.org/10.1364/OE.27.008756.
(79) Lin, Y.; Leibrandt, D. R.; Leibfried, D.; Chou, C. Quantum Entanglement between an Atom and a Molecule. *Nature* **2020**, *581* (7808), 273–277. https://doi.org/10.1038/s41586-020-2257-1.
(80) Guan, X.; Zhang, J.; Gao, X.; Wang, Y.; Shi, T.; Chen, J. A 780 Nm Optical Frequency Standard Based on Diffuse Laser Cooled 87Rb Atoms. *Appl. Phys. Lett.* **2025**, *126* (3), 031104. https://doi.org/10.1063/5.0250471.
(81) Li, C.; Chen, B.; Ruan, Z.; Wu, H.; Zhou, Y.; Liu, J.; Chen, P.; Chen, K.; Guo, C.; Liu, L. High Modulation Efficiency and Large Bandwidth Thin-Film Lithium Niobate Modulator for Visible Light. *Opt. Express* **2022**, *30* (20), 36394–36402. https://doi.org/10.1364/OE.469065.
(82) Zareen, I.; Amin, M.; Alam, M.; Ahmed, T.; Karim, M. A.; Rahman, A. Analysis of a GaAs/AlGaAs Electrooptic Modulator for High-Speed Communications. *ICECE 2010 - 6th Int. Conf. Electr. Comput. Eng.* **2010**. https://doi.org/10.1109/ICELCE.2010.5700666.
(83) Alexander, K.; George, J. P.; Verbist, J.; Neyts, K.; Kuyken, B.; Van Thourhout, D.; Beeckman, J. Nanophotonic Pockels Modulators on a Silicon Nitride Platform. *Nat. Commun.* **2018**, *9* (1), 3444. https://doi.org/10.1038/s41467-018-05846-6.
(84) Amin, R.; Maiti, R.; Carfano, C.; Ma, Z.; Tahersima, M. H.; Lilach, Y.; Ratnayake, D.; Dalir, H.; Sorger, V. J. 0.52 V Mm ITO-Based Mach-Zehnder Modulator in Silicon Photonics. *APL Photonics* **2018**, *3* (12), 126104. https://doi.org/10.1063/1.5052635.

(85) Wang, C.; Zhang, M.; Chen, X.; Bertrand, M.; Shams-Ansari, A.; Chandrasekhar, S.; Winzer, P.; Lončar, M. Integrated Lithium Niobate Electro-Optic Modulators Operating at CMOS-Compatible Voltages. *Nature* **2018**, *562* (7725), 101–104. https://doi.org/10.1038/s41586-018-0551-y.
(86) Wang, C.; Zhang, M.; Stern, B.; Lipson, M.; Lončar, M. Nanophotonic Lithium Niobate Electro-Optic Modulators. *Opt. Express* **2018**, *26* (2), 1547–1555. https://doi.org/10.1364/OE.26.001547.
(87) Abel, S.; Eltes, F.; Ortmann, J. E.; Messner, A.; Castera, P.; Wagner, T.; Urbonas, D.; Rosa, A.; Gutierrez, A. M.; Tulli, D.; Ma, P.; Baeuerle, B.; Josten, A.; Heni, W.; Caimi, D.; Czornomaz, L.; Demkov, A. A.; Leuthold, J.; Sanchis, P.; Fompeyrine, J. Large Pockels Effect in Micro- and Nanostructured Barium Titanate Integrated on Silicon. *Nat. Mater.* **2019**, *18* (1), 42–47. https://doi.org/10.1038/s41563-018-0208-0.
(88) Kamada, S.; Ueda, R.; Yamada, C.; Tanaka, K.; Yamada, T.; Otomo, A. Superiorly Low Half-Wave Voltage Electro-Optic Polymer Modulator for Visible Photonics. *Opt. Express* **2022**, *30* (11), 19771–19780. https://doi.org/10.1364/OE.456271.
(89) Renaud, D.; Assumpcao, D. R.; Joe, G.; Shams-Ansari, A.; Zhu, D.; Hu, Y.; Sinclair, N.; Loncar, M. Sub-1 Volt and High-Bandwidth Visible to near-Infrared Electro-Optic Modulators. *Nat. Commun.* **2023**, *14* (1), 1496. https://doi.org/10.1038/s41467-023-36870-w.
(90) *Mechanical properties and peculiarities of molecular crystals - Chemical Society Reviews (RSC Publishing) DOI:10.1039/D2CS00481J*. https://pubs.rsc.org/en/content/articlehtml/2023/cs/d2cs00481j (accessed 2025-03-10).
(91) Singh, B. P.; Sikarwar, S.; Agarwal, S.; Singh, D. P.; Pandey, K. K.; Manohar, R. Chemically Functionalized Gold Nanosphere-Blended Nematic Liquid Crystals for Photonic Applications. *ACS Omega* **2023**, *8* (2), 2315–2327. https://doi.org/10.1021/acsomega.2c06718.
(92) Nayek, P.; Li, G. Superior Electro-Optic Response in Multiferroic Bismuth Ferrite Nanoparticle Doped Nematic Liquid Crystal Device. *Sci. Rep.* **2015**, *5* (1), 10845. https://doi.org/10.1038/srep10845.
(93) Lee, W.; Godinho, M. H.; Yang, D.-K.; Zyryanov, V. Liquid-Crystalline Materials for Optical and Photonic Applications: Introduction to the Feature Issue. *Opt. Mater. Express* **2023**, *13* (8), 2422–2425. https://doi.org/10.1364/OME.501836.
(94) Yang, C.-S.; Lin, C.-J.; Pan, R.-P.; Que, C. T.; Yamamoto, K.; Tani, M.; Pan, C.-L. The Complex Refractive Indices of the Liquid Crystal Mixture E7 in the Terahertz Frequency Range. *JOSA B* **2010**, *27* (9), 1866–1873. https://doi.org/10.1364/JOSAB.27.001866.
(95) *Seeing the Unseen: The Role of Liquid Crystals in Gas-Sensing Technologies - Esteves - 2020 - Advanced Optical Materials - Wiley Online Library*. https://advanced.onlinelibrary.wiley.com/doi/full/10.1002/adom.201902117 (accessed 2025-03-10).
(96) Yoshizawa, A. Liquid Crystal Supermolecules Stabilizing an Optically Isotropic Phase with Frustrated Molecular Organization. *Polym. J.* **2012**, *44* (6), 490–502. https://doi.org/10.1038/pj.2012.55.

(97) Zhang, B.; Plidschun, M.; Schmidt, M. A.; Kitzerow, H.-S. Anchoring and Electro-Optic Switching of Liquid Crystals on Nano-Structured Surfaces Fabricated by Two-Photon Based Nano-Printing. *Opt. Mater. Express* **2023**, *13* (12), 3467–3480. https://doi.org/10.1364/OME.503100.
(98) Xia, Y.; Ahmed, Z.; Karimullah, A.; Mottram, N.; Heidari, H.; Ghannam, R. Thermal-Controlled Cholesteric Liquid Crystal Wavelength Filter Lens for Photosensitive Epilepsy Treatment. *Cell Rep. Phys. Sci.* **2024**, *5* (9), 102158. https://doi.org/10.1016/j.xcrp.2024.102158.
(99) Bankova, D.; Brouckaert, N.; Podoliak, N.; Beddoes, B.; White, E.; Buchnev, O.; Kaczmarek, M.; D'Alessandro, G. Characterization of Optically Thin Cells and Experimental Liquid Crystals. *Appl. Opt.* **2022**, *61* (16), 4663–4669. https://doi.org/10.1364/AO.456659.
(100) Iseghem, L. V.; Picavet, E.; Takabayashi, A. Y.; Edinger, P.; Khan, U.; Verheyen, P.; Quack, N.; Gylfason, K. B.; Buysser, K. D.; Beeckman, J.; Bogaerts, W. Low Power Optical Phase Shifter Using Liquid Crystal Actuation on a Silicon Photonics Platform. *Opt. Mater. Express* **2022**, *12* (6), 2181–2198. https://doi.org/10.1364/OME.457589.
(101) Peng, H.; Zhang, Y.; Zhu, S.; Temiz, M.; El-Makadema, A. Determining Dielectric Properties of Nematic Liquid Crystals at Microwave Frequencies Using Inverted Microstrip Lines. *Liq. Cryst.* **2022**, *49* (15), 2069–2081. https://doi.org/10.1080/02678292.2022.2102685.
(102) Jullien, A.; Bortolozzo, U.; Grabielle, S.; Huignard, J.-P.; Forget, N.; Residori, S. Continuously Tunable Femtosecond Delay-Line Based on Liquid Crystal Cells. *Opt. Express* **2016**, *24* (13), 14483–14493. https://doi.org/10.1364/OE.24.014483.
(103) Gulati, L.; Sánchez-Somolinos, C.; Giesselmann, F.; Fischer, P. Aligning and Observing the Liquid Crystal Director in 3D Using Small Magnetic Fields and a Wedge-Cell. *Adv. Funct. Mater.* **2025**, *35* (3), 2413513. https://doi.org/10.1002/adfm.202413513.
(104) Zhang, S.; Wang, Q.; Xu, B.; Hong, R.; Zhang, D.; Zhuang, S. Electrically Switchable Multicolored Filter Using Plasmonic Nanograting Integrated with Liquid Crystal for Optical Storage and Encryption. *Opt. Express* **2023**, *31* (7), 11940–11953. https://doi.org/10.1364/OE.485787.
(105) Liu, A.; Liao, L.; Rubin, D.; Nguyen, H.; Ciftcioglu, B.; Chetrit, Y.; Izhaky, N.; Paniccia, M. High-Speed Optical Modulation Based on Carrier Depletion in a Silicon Waveguide. *Opt. Express* **2007**, *15* (2), 660–668. https://doi.org/10.1364/OE.15.000660.
(106) Ozer, Y.; Kocaman, S. Stability Formulation for Integrated Opto-Mechanic Phase Shifters. *Sci. Rep.* **2018**, *8* (1), 1937. https://doi.org/10.1038/s41598-018-20405-1.
(107) *Optimisation of spontaneous four-wave mixing in a ring microcavity - IOPscience*. https://iopscience.iop.org/article/10.1070/QEL16511 (accessed 2025-03-10).
(108) Helt, L. G.; Yang, Z.; Liscidini, M.; Sipe, J. E. Spontaneous Four-Wave Mixing in Microring Resonators. *Opt. Lett.* **2010**, *35* (18), 3006–3008. https://doi.org/10.1364/OL.35.003006.
(109) Tkachenko, V.; Abbate, G.; Marino, A.; Vita, F.; Giocondo, M.; Mazzulla, A.; Ciuchi, F.; Stefano, L. D. Nematic Liquid Crystal Optical Dispersion in the Visible-Near Infrared

Range. *Mol. Cryst. Liq. Cryst.* **2006**, *454* (1), 263/[665]-271/[673]. https://doi.org/10.1080/15421400600655816.
(110) Fang, Z.; Zheng, J.; Saxena, A.; Whitehead, J.; Chen, Y.; Majumdar, A. Non-Volatile Reconfigurable Integrated Photonics Enabled by Broadband Low-Loss Phase Change Material. *Adv. Opt. Mater.* **2021**, *9* (9), 2002049. https://doi.org/10.1002/adom.202002049.
(111) Yin, Y.; Shiyanovskii, S. V.; Lavrentovich, O. D. Electric Heating Effects in Nematic Liquid Crystals. *J. Appl. Phys.* **2006**, *100* (2), 024906. https://doi.org/10.1063/1.2214466.
(112) Hinum-Wagner, J. W.; Hoermann, S. M.; Feigl, G.; Schmidt, C.; Kraft, J.; Bergmann, A. Enhancing Silicon Nitride Waveguide Performance: Optimization of Sidewall Roughness for Low-Loss Applications through Resist Reflowing via Tree-Based Modeling of Relevant Processing Parameters. In *Data Science for Photonics and Biophotonics*; SPIE, 2024; Vol. 13011, p 1301102. https://doi.org/10.1117/12.3017081.
(113) Zhao, Q.; Huang, Y.; Boyraz, O. Optical Properties of V-Groove Silicon Nitride Trench Waveguides. *JOSA A* **2016**, *33* (9), 1851–1859. https://doi.org/10.1364/JOSAA.33.001851.
(114) Basu, R.; Kehr, C. C. Accelerated Electro-Optic Switching in Liquid Crystal Devices via Ion Trapping by Dispersed Helical Carbon Nanotubes. *Micromachines* **2025**, *16* (4), 457. https://doi.org/10.3390/mi16040457.
(115) Yaroshchuk, O.; Reznikov, Y. Photoalignment of Liquid Crystals: Basics and Current Trends. *J. Mater. Chem.* **2011**, *22* (2), 286–300. https://doi.org/10.1039/C1JM13485J.
(116) Dutta, J.; Chen, R.; Tara, V.; Majumdar, A. Low-Power Quasi-Continuous Hybrid Volatile/Nonvolatile Tuning of Ring Resonators. **2025**.
(117) Wang, J.; Liu, K.; Harrington, M. W.; Rudy, R. Q.; Blumenthal, D. J. Silicon Nitride Stress-Optic Microresonator Modulator for Optical Control Applications. *Opt. Express* **2022**, *30* (18), 31816–31827. https://doi.org/10.1364/OE.467721.
(118) Deckoff-Jones, S.; Wang, Y.; Lin, H.; Wu, W.; Hu, J. 2D-Material-Enabled Multifunctional Mid-IR Optoelectronics. In *Smart Photonic and Optoelectronic Integrated Circuits XXII*; He, S., Vivien, L., Eds.; SPIE: San Francisco, United States, 2020; p 64. https://doi.org/10.1117/12.2543619.
(119) Alexander, K.; George, J. P.; Verbist, J.; Neyts, K.; Kuyken, B.; Van Thourhout, D.; Beeckman, J. Nanophotonic Pockels Modulators on a Silicon Nitride Platform. *Nat. Commun.* **2018**, *9* (1), 3444. https://doi.org/10.1038/s41467-018-05846-6.
(120) Xu, M.; He, M.; Zhang, H.; Jian, J.; Pan, Y.; Liu, X.; Chen, L.; Meng, X.; Chen, H.; Li, Z.; Xiao, X.; Yu, S.; Yu, S.; Cai, X. High-Performance Coherent Optical Modulators Based on Thin-Film Lithium Niobate Platform. *Nat. Commun.* **2020**, *11* (1), 3911. https://doi.org/10.1038/s41467-020-17806-0.
(121) Tamalampudi, S. R.; Dushaq, G. H.; Villegas, J. E.; Paredes, B.; Rasras, M. S. A Multi-Layered GaGeTe Electro-Optic Device Integrated in Silicon Photonics. *J. Light. Technol.* **2023**, *41* (9), 2785–2791. https://doi.org/10.1109/JLT.2023.3237818.
(122) Liu, G.; Yu, H.; Ban, D.; Li, B.; Wei, G.; Yang, C.; Wang, J.; Sohn, Y.-I.; Han, Y.; Qiu, F. Highly Efficient Lead Zirconate Titanate Ring Modulator. *APL Photonics* **2024**, *9* (6), 066111. https://doi.org/10.1063/5.0193922.

(123) Zhou, W.; Zhang, Y.; Jiang, Y.; Zhang, P.; Shen, J.; Zhang, X.; Chen, Y.; Sun, M.; Qiu, F.; Xiao, H.; Tian, Y.; Su, Y. Integrated Electro-Optic Modulator on a Lead Zirconate Titanate-Silicon Nitride Heterogeneous Platform. *Opt. Lett.* **2024**, *49* (22), 6353–6356. https://doi.org/10.1364/OL.538022.

(124) Deckoff-Jones, S.; Wang, Y.; Lin, H.; Wu, W.; Hu, J. Tellurene: A Multifunctional Material for Midinfrared Optoelectronics. *ACS Photonics* **2019**, *6* (7), 1632–1638. https://doi.org/10.1021/acsphotonics.9b00694.

(125) Eltes, F.; Mai, C.; Caimi, D.; Kroh, M.; Popoff, Y.; Winzer, G.; Petousi, D.; Lischke, S.; Ortmann, J. E.; Czornomaz, L.; Zimmermann, L.; Fompeyrine, J.; Abel, S. A $BaTiO_3$ - Based Electro-Optic Pockels Modulator Monolithically Integrated on an Advanced Silicon Photonics Platform. *J. Light. Technol.* **2019**, *37* (5), 1456–1462. https://doi.org/10.1109/JLT.2019.2893500.

(126) Barzic, A. I.; Ioan, S.; Barzic, A. I.; Ioan, S. Viscoelastic Behavior of Liquid-Crystal Polymer in Composite Systems. In *Viscoelastic and Viscoplastic Materials*; IntechOpen, 2016. https://doi.org/10.5772/64074.

(127) Gennes, P. G. D.; Prost, J. Dynamical Properties Of Nematics. In *The Physics of Liquid Crystals*; Gennes, P. G. D., Prost, J., Eds.; Oxford University Press, 1993; p 0. https://doi.org/10.1093/oso/9780198520245.003.0005.

(128) Gennes, P. G. D.; Prost, J. Liquid Crystals: Main Types and Properties. In *The Physics of Liquid Crystals*; Gennes, P. G. D., Prost, J., Eds.; Oxford University Press, 1993; p 0. https://doi.org/10.1093/oso/9780198520245.003.0001.

(129) Pasechnik, S. V.; Chigrinov, V. G.; Shmeliova, D. V. Liquid Crystals: Viscous and Elastic Properties.

(130) Zakharov, A. V.; Mirantsev, L. V. Dynamic and Dielectric Properties of Liquid Crystals. *Phys. Solid State* **2003**, *45* (1), 183–188. https://doi.org/10.1134/1.1537433.

(131) Williams, G. Dielectric Relaxation Behaviour of Liquid Crystals. In *The Molecular Dynamics of Liquid Crystals*; Luckhurst, G. R., Veracini, C. A., Eds.; Springer Netherlands: Dordrecht, 1994; pp 431–450. https://doi.org/10.1007/978-94-011-1168-3_17.

(132) Hobbs, J.; Reynolds, M.; Krishnappa Srinatha, M.; Shanker, G.; Mattsson, J.; Nagaraj, M. The Relaxation Dynamics and Dielectric Properties of Cyanobiphenyl-Based Nematic Tripod Liquid Crystals. *J. Mol. Liq.* **2023**, *391*, 123069. https://doi.org/10.1016/j.molliq.2023.123069.

(133) Jonscher, A. K. The Physical Origin of Negative Capacitance. *J. Chem. Soc. Faraday Trans. 2 Mol. Chem. Phys.* **1986**, *82* (1), 75–81. https://doi.org/10.1039/F29868200075.

(134) Chen, H.-Y.; Yang, K.-X.; Lee, W. Transient Behavior of the Polarity-Reversal Current in a Nematic Liquid-Crystal Device. *Opt. Express* **2004**, *12* (16), 3806–3813. https://doi.org/10.1364/OPEX.12.003806.

(135) Sugimura, A.; Matsui, N.; Takahashi, Y.; Sonomura, H.; Naito, H.; Okuda, M. Transient Currents in Nematic Liquid Crystals. *Phys. Rev. B* **1991**, *43* (10), 8272–8276. https://doi.org/10.1103/PhysRevB.43.8272.

(136) Jia, X.; Zhai, C.; Zhu, X.; You, C.; Cao, Y.; Zhang, X.; Zheng, Y.; Fu, Z.; Mao, J.; Dai, T.; Chang, L.; Su, X.; Gong, Q.; Wang, J. Continuous-Variable Multipartite Entanglement in

an Integrated Microcomb. *Nature* **2025**, *639* (8054), 329–336. https://doi.org/10.1038/s41586-025-08602-1.

(137) Smith, B. J.; Kundys, D.; Thomas-Peter, N.; Smith, P. G. R.; Walmsley, I. A. Phase-Controlled Integrated Photonic Quantum Circuits. *Opt. Express* **2009**, *17* (16), 13516–13525. https://doi.org/10.1364/OE.17.013516.

(138) Van Iseghem, L.; Khan, U.; Picavet, E.; Takabayashi, A. Y.; Edinger, P.; Verheyen, P.; Quack, N.; Gylfason, K. B.; De Buysser, K.; Beeckman, J.; Bogaerts, W. Efficient Low-Voltage Phase Shifter with Inkjet-Printed Liquid Crystal on a Silicon on Insulator Platform. *Adv. Opt. Mater. n/a* (n/a), 2500186. https://doi.org/10.1002/adom.202500186.

(139) Zhang, W.; Pivnenko, M.; Chu, D. Fast-Speed Electrically Tunable Liquid Crystal Metasurface for 2π Phase Modulation at Normal and Oblique Angles of Incidence. *Opt. Express* **2025**, *33* (4), 8911–8923. https://doi.org/10.1364/OE.554542.

(140) Wang, X.; Grist, S.; Flueckiger, J.; Jaeger, N. A. F.; Chrostowski, L. Silicon Photonic Slot Waveguide Bragg Gratings and Resonators. *Opt. Express* **2013**, *21* (16), 19029–19039. https://doi.org/10.1364/OE.21.019029.

(141) De Cort, W.; Beeckman, J.; Claes, T.; Neyts, K.; Baets, R. Wide Tuning of Silicon-on-Insulator Ring Resonators with a Liquid Crystal Cladding. *Opt. Lett.* **2011**, *36* (19), 3876. https://doi.org/10.1364/OL.36.003876.

**Supplementary Information for Near-visible low power tuning of nematic-liquid crystal integrated silicon nitride ring resonator**

Jayita Dutta, [*,†] Antonio Ferraro, [§] Arnab Manna,[¶] Rui Chen,[†] Alfredo Pane[§], Giuseppe Emanuele Lio,[||] Roberto Caputo,[‡, §] and Arka Majumdar[*,†,¶]

† Electrical and Computer Engineering, University of Washington, Seattle, WA, 98195, USA.

,§ Consiglio Nazionale delle Ricerche - Istituto di Nanotecnologia CNR-Nanotec, Rende (CS), 87036 Italy

¶ Department of Physics, University of Washington, Seattle, WA, 98195, USA.

|| Istituto di Nanoscienze CNR-NANO, Consiglio Nazionale delle Ricerche, Pisa, 56127, Italy

‡ University of Calabria, I-87036 Rende (CS), Italy.

E-mail: jayitad@uw.edu; arka@uw.edu

**Section 1. Lumerical Mode Simulation to study the impact of buffer $SiO_2$ layer between the liquid crystal (LC) and silicon nitride (SiN) waveguides**

We studied the effect of a buffer $SiO_2$ layer between the liquid crystal (LC) and silicon nitride (SiN) waveguides for metal waveguide separation gap of $1\mu m$ and the same is shown in Supplementary Figure 1.

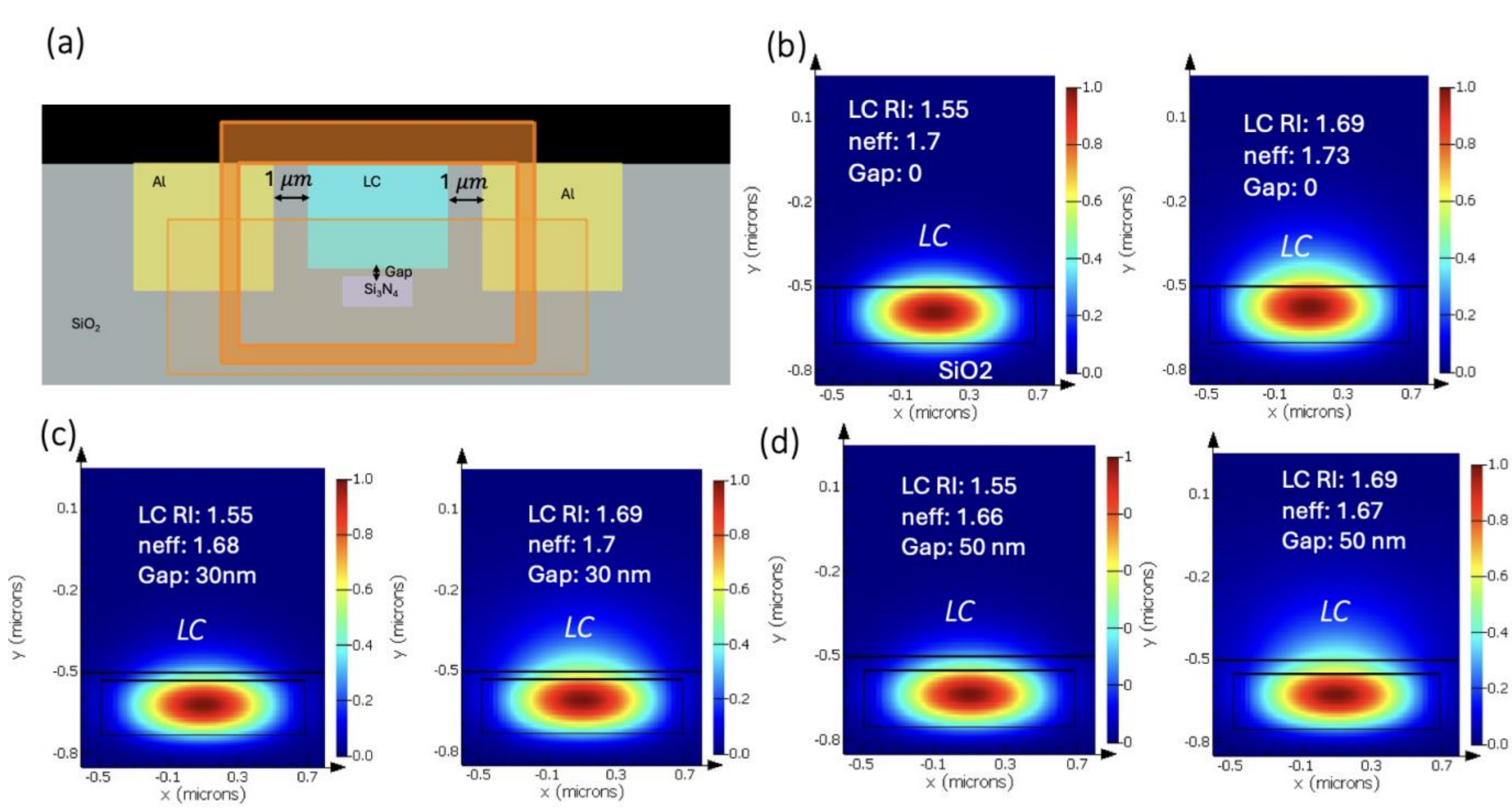

Supplementary Figure 1 : Lumerical Mode Simulation to study the impact of buffer SiO2 layer between the LC and SiN waveguides for metal waveguide separation gap of $1\mu$m. **a.** Cross sectional view of the device with buffer SiO2 layer. Variation in effective mode index for buffer SiO2 layer of **b.** 0, **c.** 30 nm, **d.** 50 nm

As the buffer $SiO_2$ layer increased from 0 to 50 nm (represented by the Gap in Supplementary Figure 1a), a decrease in effective mode index is observed w.r.t. to the LC refractive index. When there is no buffer $SiO_2$ layer (i.e. Gap = 0 , Supplementary Figure 1b) between the waveguide and LC trench, the change in effective mode index ($\Delta n_{eff} = 0.03$) is maximum which in turn will maximize the induced phase shift for a given length of the phase shifter. The reason for this is the interaction between the waveguide mode and LC region increases if there is no oxide in between, which, therefore, results in a greater variation in the effective index of the waveguide mode. The change in effective mode index decreases to 0.02 (Supplementary Figure 1c) and 0.01 (Supplementary Figure 1d) for an increase in buffer oxide layer to 30nm and 50 nm respectively.

**Section 2. Optical and Electrical Characterization Setup**

A glimpse of the electrical and optical characterization setup is shown in Supplementary Figure 3a. Input light is provided by the laser, and optical fibers are coupled to the on-chip gratings. The fabricated chip with LC integrated SiN ring resonators is shown in Supplementary Figure 3d. A picture of the LC integrated SiN ring resonator with metal waveguide separation gap of 1$\mu$m is shown in Supplementary Figure 3b. For on-chip external electric field, electrical pulses were applied to the on-chip metal contacts via a pair of electrical probes as shown in Supplementary Figure 3c.

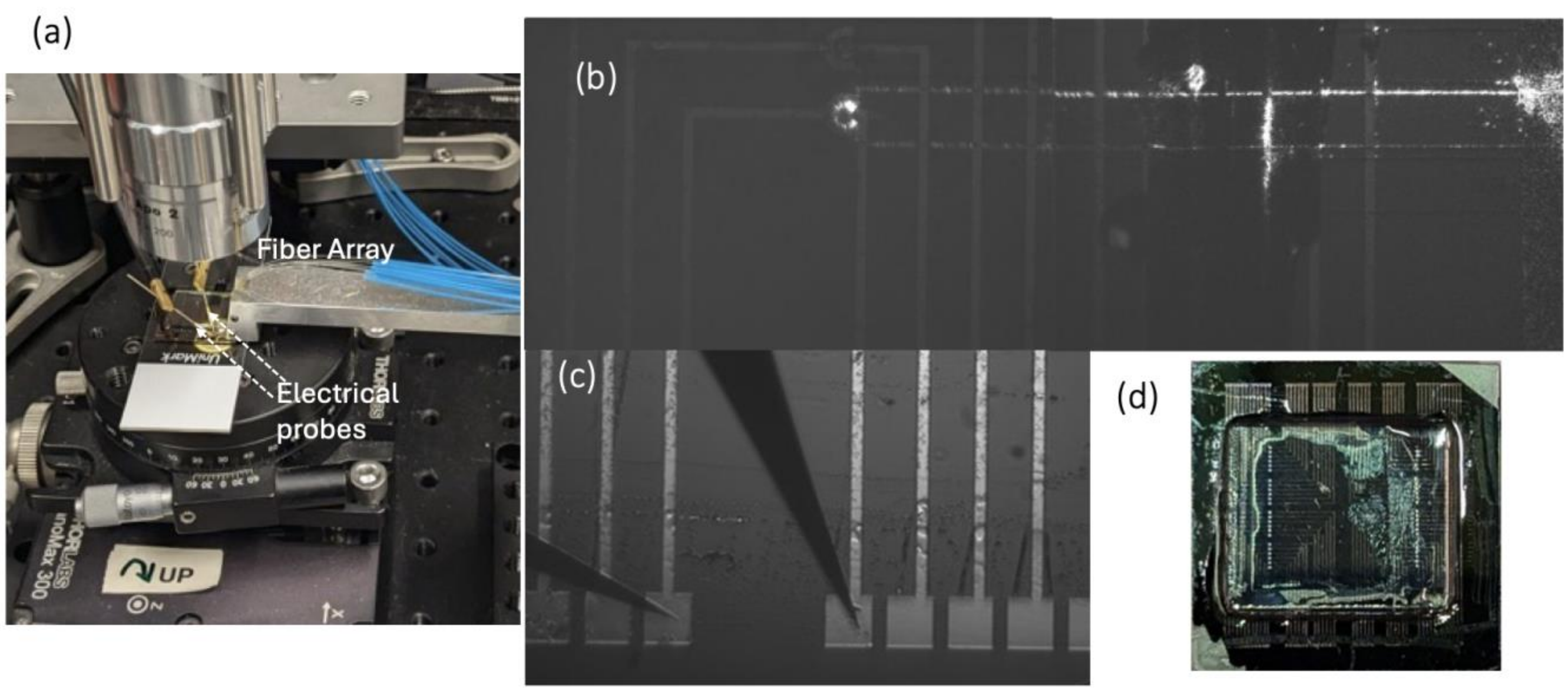

Supplementary Figure 2: **a.** Optical and Electrical Characterization Setup **(b-c).** SiN ring resonator with metal waveguide separation gap of 1μm **d.** The fabricated chip with liquid crystal integrated in SiN platform